\documentclass{LMCS}

\def\Rlap#1{\rlap{$#1$}}
\def\Llap#1{\llap{$#1$}}

\usepackage{ifthen}
\newboolean{long}
\setboolean{long}{false}
\usepackage{url}

\usepackage{color}
\usepackage{amssymb}

\usepackage{amsmath}
\usepackage{amsfonts}
\usepackage{enumerate}
\usepackage{proof}
\usepackage{611}
\usepackage{hyperref}

\newtheorem{definition}[thm]{Definition}

\newtheorem{corollary}[thm]{Corollary}

\newtheorem{lemma}[thm]{Lemma}

\newcommand\iffl{\ensuremath{\leftrightarrow}}

\renewcommand\int{\mathbf{Z}}

\newcommand{\ignore}[1]{}
\newcommand{\todo}[1]{}
\newcommand{\ov}[1]{\ensuremath{\overline{#1}}}
\newcommand{\nat}{\ensuremath{\mathbb{N}}}

\newcommand\reals{\ensuremath{\Vdash}}
\newcommand{\p}{\proves}

\newcommand{\g}{\Gamma}

\newcommand{\gp}{\Gamma \proves}

\newcommand{\og}{\ov{\g}}
\newcommand{\rrho}{\reals_\rho}

\newcommand{\pl}[1]{\ensuremath{\mathrm{#1}}}
\newcommand{\FST}{\pl{fst}}
\newcommand{\SND}{\pl{snd}}
\newcommand{\LET}{\pl{let}}
\newcommand{\CASE}{\pl{case}}\newcommand{\DOM}{\pl{dom}}
\newcommand{\MAGIC}{\pl{magic}}
\newcommand{\INL}{\pl{inl}}
\newcommand{\INR}{\pl{inr}}
\newcommand{\IN}{\pl{in}}
\newcommand{\IND}{\pl{ind}}
\newcommand{\sr}[1]{\ensuremath{\SB{#1}_\rho}}
\newcommand{\izfc}{IZF${}_C$}
\newcommand{\izfr}{IZF${}_R$}
\newcommand{\iizf}{IZF${}_R^{-}$}
\newcommand{\iizfr}{IZF${}_R^{-}$}
\newcommand{\li}{\lambda Z}
\newcommand{\la}{\lambda Z}
\newcommand{\lz}{\lambda Z}
\newcommand{\lrk}{\ensuremath{\lambda rk}}
\def\squareforqed{\hbox{\rlap{$\sqcap$}$\sqcup$}}

\def\doi{4 (2:1) 2008}
\lmcsheading%
{\doi}
{1--29}
{}
{}
{Jan.~\phantom{0}2, 2007}
{Apr.~08, 2008}
{}   

\begin{document}

\title{Normalization of IZF with Replacement}

\author{Wojciech Moczyd\l owski}
\address{Department of Computer Science, Cornell University, Ithaca, NY\ 14853, USA}
\email{wojtek@cs.cornell.edu}
\thanks{Partly supported by NSF grants DUE-0333526 and 0430161.}

\titlecomment{}

\begin{abstract}
\noindent
IZF is a well investigated impredicative constructive version of
Zermelo-Fraen\-kel set theory. Using set terms, we axiomatize IZF with
Replacement, which we call \izfr, along with its intensional
counterpart \iizfr. We define a typed lambda calculus $\li$
corresponding to proofs in \iizfr\ according to the Curry-Howard
isomorphism principle. Using realizability for \iizfr, we show weak
normalization of $\li$. We use normalization to prove the disjunction,
numerical existence and term existence properties.  An inner
extensional model is used to show these properties, along with the set
existence property, for full, extensional \izfr.
\end{abstract}

\maketitle

\section{Introduction}

Four salient properties of constructive set theories are:
\begin{enumerate}[$\bullet$]
\item Numerical Existence Property (NEP): From a proof of a statement
``there exists a natural number $x$ such that {\ldots}'' a witness $n \in
\nat$ can be extracted. 
\item Disjunction Property (DP): If $\phi \lor \psi$ is provable, then
either $\phi$ or $\psi$ is provable.
\item Term Existence Property (TEP): If $\exists x.\ \phi(x)$ is
provable, then $\phi(t)$ is provable for some term $t$.
\item Set Existence Property (SEP): If $\exists x.\ \phi(x)$ is
provable, then there is a formula $\psi(x)$ such that $\exists! x.\ \phi(x)
\land \psi(x)$ is provable, where both $\phi$ and $\psi$ are term-free. 
\end{enumerate}

How to prove these properties for a given theory? There is a variety of
methods applicable to constructive theories. Cut-elimination, proof
normalization, realizability, Kripke models{\ldots}.
Normalization proofs, based on the Curry-Howard isomorphism principle, have the advantage of
providing an explicit method of witness and program extraction from
proofs. They also provide information about the behaviour of the proof
system. 

We are interested in intuitionistic set theory IZF. It is essentially what 
remains of ZF set theory after excluded middle is carefully taken away. An important 
decision to make on the way is whether to use Replacement or Collection axiom schema. 
We will call the version with Collection \izfc\  and the version with Replacement \izfr. In the literature,
IZF usually denotes \izfc. Both theories extended with excluded middle are
equivalent to ZF \cite{friedmancons}. They are not equivalent \cite{frsce3}.
While the proof-theoretic power of \izfc\ is equivalent to that of ZF, the exact
power of \izfr\ is unknown. Arguably \izfc\ is less constructive, as
Collection, similarly to Choice, asserts the existence of a set without
defining it. 

Both versions have been investigated thoroughly. Results up to 1985 are
presented in \cite{beesonbook,scedrov85}. Later research was concentrated on
weaker subsystems \cite{ar,ikp}. A predicative constructive set theory CZF has attracted
particular interest. \cite{ar} describes the set-theoretic apparatus
available in CZF and provides further references.

We axiomatize \izfr, along with its intensional version \iizfr, using set
terms. We define a typed lambda calculus $\li$ corresponding to proofs in
\iizfr. We also define realizability for \iizfr, in the spirit of
\cite{mccarty}, and use it to show that $\li$ weakly normalizes. Strong normalization of $\li$ does not hold; moreover, we show
that in non-well-founded IZF even weak normalization fails.

With normalization in hand, the properties NEP, DP and TEP
easily follow. To show these properties for full, extensional \izfr, we define an inner
model $T$ of \izfr, consisting of what we call transitively L-stable sets.
We show that a formula is true in \izfr\ iff its relativization to $T$ is true
in \iizfr. Therefore \izfr\ is interpretable in \iizfr. This allows us to use 
the properties proven for \iizfr. In \izfr, SEP easily follows from TEP. 

The importance of these properties in the context of computer science stems
from the fact that they make it possible to extract programs from
constructive proofs. For example, suppose \izfr\ $\p \forall n \in \nat
\exists m \in \nat.\ \phi(n, m)$. From this proof a program can be extracted --- 
take a natural number $n$, construct a proof \izfr\ $\p \ov{n}\in \nat$. Combine the proofs to get \izfr\ $ \p \exists m \in \nat.\ \phi(\ov{n}, m)$ and apply 
NEP to get a number $m$ such that \izfr\ $\p \phi(\ov{n}, \ov{m})$. A
detailed account of program extraction from \izfr\ proofs can be found in \cite{chol}.

There are many provers with the program extraction capability. However, 
they are usually based on variants of type theory, which is a foundational basis very different from set
theory. This makes the process of formalizing program specification more
difficult, as an unfamiliar new language and logic have to be learned from
scratch. \cite{lamport99} strongly argues \emph{against} using type theory
for the specification purposes, instead promoting standard set theory. 

\izfr\ provides therefore the best of both worlds. It is a set theory,
with familiar language and axioms. At the same time, programs can be
extracted from proofs. Our $\li$ calculus and the normalization
theorem make the task of constructing the prover based on \izfr\ not very
difficult.

This paper is mostly self-contained. We assume some familiarity with
set theory, proof theory and programming languages terminology, found for example 
in \cite{kunen,urzy,pierce}. The paper is organized as follows. We start by presenting in details
intuitionistic first-order logic in section \ref{ifol}. In section \ref{izf} we
define \izfr\ along with its intensional version \iizfr. In section \ref{lz} we define a
lambda calculus $\li$ corresponding to \iizfr\ proofs. Realizability for
\iizfr\ is defined in section \ref{izfreal}. We use it to prove normalization of $\li$ in
section \ref{sectionnorm}, where we also show that non-well-founded IZF does
not normalize. We prove the properties in section \ref{secapp}, and show how to derive them 
for full, extensional \izfr\ in section \ref{lei}. Comparison with other results can be found in section \ref{others}.

\section{Intuitionistic first-order logic}\label{ifol}

Due to the syntactic character of our results, we present the intuitionistic first-order logic
(IFOL) in details. We use a natural deduction style of proof rules. The terms will be denoted by
letters $t, s, u$. The variables will be denoted by letters $a, b, c, d, e,
f$. The notation $\vec{a}$ stands for a finite sequence, treated as a set when
convenient. The $i$-th element of a sequence is denoted by $a_i$. We consider $\alpha$-equivalent
formulas equal. The capture-avoiding substitution is defined as usual; the
result of substituting $s$ for $a$ in a term $t$ is denoted by $t[a:=s]$. We
write $t[a_1, {\ldots} , a_n := s_1, {\ldots} , s_n]$ to denote the result
of substituting simultaneously $s_1, {\ldots} , s_n$ for $a_1, {\ldots} ,
a_n$. Contexts, denoted by $\Gamma$, are sets of formulas. 
The set of free variables of a formula $\phi$, denoted by $FV(\phi)$, are
defined as usual. The free variables of a context $\g$, denoted by $FV(\g)$, are
the free variables of all formulas in $\g$. The notation $\phi(\vec{a})$ means
that all free variables of $\phi$ are among $\vec{a}$. The proof rules are as follows:
\[
\infer{\g, \phi \p \phi}{} \qquad \infer{\gp \phi}{\gp \bot} 
\qquad
\infer{\gp \psi}{\gp \phi \to \psi & \gp \phi} \qquad \infer{\gp \phi \to \psi}{\g, \phi \p \psi}
\]
\[
\infer{\gp \phi \land \psi}{\gp \phi & \gp \psi} \qquad \infer{\gp
\phi}{\gp \phi \land \psi} \qquad \infer{\gp \psi}{\gp \phi \land \psi}
\]
\[
\infer{\gp \phi \lor \psi}{\gp \phi} \quad \infer{\gp \phi \lor
\psi}{\gp \psi} \quad
\infer{\gp \vartheta}{\gp \phi \lor \psi & \g, \phi \proves \vartheta & \g, \psi
\proves \vartheta}
\]
\[
\infer[a \notin FV(\g)]{\gp \forall a.\ \phi}{\gp  \phi} \qquad
\infer{\gp \phi[a:=t]}{\gp \forall a.\ \phi}
\]
\[
\infer{\gp \exists a.\ \phi}{\gp \phi[a:=t]} \qquad \infer[a \notin
FV(\g) \cup \{ \psi \}]{\gp \psi}{\gp \exists a.\ \phi &
\g, \phi \p \psi}
\]

Negation in IFOL is an abbreviation: $\lnot \phi \equiv \phi \to
\bot$. So is the symbol $\iffl$: $\phi \iffl \psi \equiv (\phi \to \psi \land
\psi \to \phi)$. Note that IFOL does not contain equality. The excluded middle rule added to IFOL makes it equivalent
to the classical first-order logic without equality. We adopt the
``dot''-convention --- a formula $\forall a.\ \phi$ should be parsed as
$\forall a.\ (\phi)$. In other words\footnote{Borrowed from \cite{urzy}.}, the
dot represents a left parenthesis whose scope extends as far to the right as
possible. 


\begin{lemma}\label{formsubst}
For any formula $\phi$, $\phi[a:=t][b:=u[a:=t]] = \phi[b:=u][a:=t]$, for $b \notin FV(t)$. 
\end{lemma}
\proof
Straightforward structural induction on $\phi$.\qed

\section{\izfr}\label{izf}

Intuitionistic set theory \izfr\ is a first-order theory, equivalent to ZF
when extended with excluded middle. It is a definitional extension of
term-free versions presented in \cite{myhill72,beesonbook,frsce3}.
The signature consists of one binary relational symbol $\in$ and function symbols used in the axioms below.
The set of all \izfr\ terms will be denoted by $Tms$. The notation $t = u$ is an abbreviation for $\forall z.\ z \in t
\iffl z \in u$. Function symbols $0$ and $S(t)$ are abbreviations for
$\emptyset$ and $\bigcup \{ t, \{ t, t \} \}$. Bounded quantifiers and the quantifier $\exists !a$ (there exists exactly one $a$) are also abbreviations 
defined in the standard way. The axioms are as follows:

\begin{enumerate}[$\bullet$]
\item (EMPTY) $\forall c.\ c \in \emptyset \iffl \bot$
\item (PAIR) $\forall a, b \forall c.\ c \in \{ a, b \} \iffl c = a \lor
c = b$
\item (INF) $\forall c.\ c \in \omega \iffl c = 0 \lor \exists b \in \omega.\ c =
S(b)$
\item (SEP${}_{\phi(a, \vec{f})}$)
$\forall \vec{f}, a \forall
c.\ c \in S_{\phi(a, \vec{f})}(a, \vec{f}) \iffl c \in a \land
\phi(c, \vec{f})$
\item (UNION) $\forall a \forall c.\ c \in \bigcup a \iffl \exists b \in
a.\ c \in b$
\item (POWER) $\forall a \forall c.\ c \in P(a) \iffl \forall b.\ b \in c \to b \in a$
\item (REPL${}_{\phi(a, b, \vec{f})}$) $\forall \vec{f}, a
\forall c.\ c \in R_{\phi(a, b, \vec{f})}(a, \vec{f}) \iffl
(\forall x \in a \exists! y.\ \phi(x, y, \vec{f})) \land (\exists x \in
a.\ \phi(x, c, \vec{f}))$
\item (IND${}_{\phi(a, \vec{f})}$) $\forall \vec{f}.\ (\forall a.\ 
(\forall b \in a.\ \phi(b, \vec{f})) \to \phi(a, \vec{f})) \to \forall a.\ \phi(a, \vec{f})$
\item (L${}_{\phi(a, \vec{f})}$) $ \forall \vec{f}, a, b.\ a = b \to \phi(a, \vec{f}) \to \phi(b, \vec{f})$
\end{enumerate}

Axioms SEP${}_\phi$, REPL${}_\phi$, IND${}_\phi$ and L${}_\phi$ are axiom schemas, and
so are the corresponding function symbols ---
there is one function symbol for each formula $\phi$. Formally, we define formulas and terms by mutual induction:
\[
\phi ::= t \in t\ |\ \phi \land \phi\ | {\ldots} \qquad \qquad t ::= a\ |\ \{ t, t \}\ |\ \ S_{\phi(a, \vec{f})}(t, \vec{t})\ |\ R_{\phi(a,
b, \vec{f})}(t, \vec{t})\ | {\ldots} 
\]

Our presentation is not minimal; for example, the empty set axiom can be
derived as usual using Separation and Infinity. However, we aim for a
\emph{natural} axiomatization of \izfr, not necessarily the most optimal one. 

The Leibniz axiom schema L${}_\phi$ is usually not present among the axioms of set
theories, as it is assumed that logic contains equality and the axiom is 
a proof rule. We include L${}_\phi$ among the axioms of \izfr, because
there is no obvious way to add it to intuitionistic logic in the Curry-Howard isomorphism context,
as its computational content is unclear. Our axiom of Replacement is
equivalent to the usual formulations, see \cite{jatrinac2006} for details.

\iizfr\ will denote \izfr\  without the Leibniz axiom schema L${}_\phi$. \iizfr\ is
an intensional version of \izfr\  --- even though extensional equality is used
in the axioms, it does not behave as the ``real'' equality. 

The terms $S_\phi(a, \vec{f})$ and $R_\phi(a, \vec{f})$ can be displayed
as $\{ x \in a\ |\ \phi(x, \vec{f}) \}$ and $\ \{ z\ |\ (\forall x \in a
\exists !y.\ \phi(x, y, \vec{f})) \land \exists x \in a.\ \phi(x, z, \vec{f}) \}$.

The axioms (EMPTY), (PAIR), (INF), (SEP${}_{\phi}$), (UNION), (POWER) and (REPL$_{\phi}$)
all assert the existence of certain classes and have the same form: $\forall 
\vec{a}. \forall c.\ c \in t_A(\vec{a}) \iffl \phi_A(c, \vec{a})$, where $t_A$ is a 
function symbol and $\phi_A$ a corresponding formula
for the axiom A. For example, for (POWER), $t_{\mathit{POWER}}$ is $P$ and
$\phi_{\mathit{POWER}}$ is $\forall b.\ b \in c
\to b \in a$. We reserve the notation $t_A$ and $\phi_A$ to denote the term and
the corresponding formula for the axiom A.

\begin{lemma}\label{tdef0}
Every term $T \equiv t_A(\overrightarrow{t(\vec{a})})$ of \izfr\ is definable. In
other words, there is a term-free formula $\phi(x,\vec{a})$ such that \izfr 
$\p \forall \vec{a}.\ \phi(T, \vec{a}) \land \exists !x.\ \phi(x, \vec{a})$.
\end{lemma}
\proof
Straightforward induction on the size of $T$.
We first show the claim for $\omega$, then for the rest of the terms. For $\omega$, the defining
formula\footnote{Strictly speaking, it is not term-free, but eliminating
terms used in $\phi$ is straightforward.} is:
\[
\phi(x) \equiv c \in x \iffl c = 0 \lor \exists y \in x.\ c = S(y)
\]
Indeed, $\phi(\omega)$ holds. Suppose $\phi(z)$ for some
$z$, we need to show that $z = \omega$. To do this, we prove by
$\in$-induction $\forall c.\ c \in z \iffl c \in \omega$. Take any $c$ and 
suppose $c \in z$. Then $c = 0$ or there is $y \in z$ such that $c = S(y)$.
In the former case $c \in \omega$, in the latter $y \in c$, so by the
induction hypothesis $y \in \omega$ and hence $c \in \omega$. The other
direction is symmetric. 

Consider now arbitrary $T \equiv t_A(\overrightarrow{t(\vec{a})})$. Let
$\vec{u}$ denote $\overrightarrow{t(\vec{a})}$, so $T \equiv t_A(\vec{u})$.
By the induction hypothesis there are formulas $\overrightarrow{\phi(x,
\vec{a})}$ defining $\vec{u}$. Consider the formula:
\[
\phi(x, \vec{a}) \equiv \exists \vec{x}.\ \bigwedge \overrightarrow{\phi(x,
\vec{a})} \land \forall c.\ c \in x \iffl \phi_A(c, \vec{x})
\]
We will now show that $\phi(x, \vec{a})$ defines $T$. Take any $\vec{a}$ and take 
$\vec{x} = \vec{u}$. We have
$\bigwedge \overrightarrow{\phi(u, \vec{a})}$ and
by the axiom (A) corresponding to $t_A$, we get $\forall c.\ c \in
t_A(\vec{u}) \iffl \phi_A(c, \vec{u})$. Furthermore, suppose $\phi(z, \vec{a})$ for
some $z$. Then there are $\vec{b}$ such that $\bigwedge
\overrightarrow{\phi(b, \vec{a})}$ and $\forall c.\
c \in z \iffl \phi_A(c, \vec{b})$. Since $\overrightarrow{\phi(x, \vec{a})}$ define $\vec{u}$,
$\vec{b} = \vec{u}$ and thus also $\forall c.\ c \in z \iffl \phi_A(c, \vec{u})$.
To show that $z = T$, it suffices to show that $\forall a.\ a \in T \iffl a \in z$, which follows easily.

It remains to consider the situation when $\phi_A$ contains some terms,
which can happen if $A$ is the Separation or Replacement axiom. However, by the 
induction hypothesis all these terms are definable as well, so there is also a term-free formula $\phi'$ equivalent to $\phi$.
\ignore{
\[
\phi(x, \vec{a}) \equiv \exists \vec{x}.\ \bigwedge \overrightarrow{\phi(x,
\vec{a})} \land \forall c.\ c \in x \iffl \phi_A(c, \vec{x})
\]
Let the formulas
$\overrightarrow{\phi(x, \vec{a})}$ define $\overrightarrow{u(\vec{a})}$.
defines $t$. To see that $\phi(t)$ holds, take $\vec{x} = \vec{u}$.
By the induction hypothesis we have $\bigwedge \overrightarrow{\phi(u)}$ and 
by the axiom (A) defining $t_A$, we get $\forall c.\ c \in
t_A(\vec{u}) \iffl \phi_A(c, \vec{u})$. Furthermore, suppose $\phi(z)$ for
some $z$. Then there are $\vec{b}$ such that $\bigwedge \overrightarrow{\phi(b)}$ and $\forall c.\
c \in z \iffl \phi_A(c, \vec{b})$. Since $\overrightarrow{\phi(x)}$ define $\vec{u}$,
$\vec{b} = \vec{u}$ and thus also $\forall c.\ c \in z \iffl \phi_A(c, \vec{u})$. 
To show that $z = t$, it suffices to show that $\forall a.\ a \in t \iffl a
\in z$, which follows easily.}\qed

\begin{corollary}\label{tdef}
For any closed term $t$ there is a term-free formula $\phi(x)$ such that \izfr $\p 
(\exists !x.\ \phi(x)) \land \phi(t)$.
\end{corollary}

\section{The $\li$ calculus}\label{lz}

We now present a lambda calculus $\li$ for \iizfr, based on the Curry-Howard isomorphism
principle. The first-order part of $\li$ is essentially $\lambda P1$ from
\cite{urzy}. The lambda terms in the calculus correspond to proofs in \iizfr.
The correspondence is captured formally by Lemma \ref{ch}. 

The lambda terms in $\li$ will be denoted by letters $M, N, O, P$. 
There are two kinds of lambda abstractions, one used for proofs of implications, the other for proofs of
universal quantifications. We use separate sets of variables for these abstractions and call them
proof and first-order variables, respectively. We use letters $x, y,
z$ for proof variables and $a, b, c$ for first-order variables. 
Letters $t, s, u$ are reserved for \izfr\ terms. The types in the system are
\izfr\ formulas. The lambda terms are generated by an abstract grammar. The
first group of terms is standard and used for IFOL proofs:
\begin{eqnarray*}
M & ::= & x\ |\ M\ N\ |\ \lambda a.\ M\ |\ \lambda x : \phi.\ M\ |\ \INL(M)\ |\
\INR(M) \ |\ \FST(M)\ | \ \SND(M)\ |\ [t, M]\ |\ M\ t\\
& & <M, N> \ |\ \CASE(M, x : \phi.\ N, x : \psi.\ O)\ |\ \MAGIC(M)\ |\ \LET\
[a, x : \phi] := M\ \IN\ N
\end{eqnarray*}
The rest of the terms correspond to the axioms of \iizfr:
\[
\pl{emptyProp}(t, M)\ |\ \pl{emptyRep}(t, M)
\]
\[
\pl{pairProp}(t, u_1, u_2, M)\ |\ \pl{pairRep}(t, u_1, u_2, M)
\]
\[
\pl{unionProp}(t, u, M)\ |\ \pl{unionRep}(t, u, M)
\]
\[
\pl{sep}_{\phi(a, \vec{f})}\pl{Prop}(t, u, \vec{u}, M)\ |\ \pl{sep}_{\phi(a, \vec{f})}\pl{Rep}(t, u, \vec{u}, M)
\]
\[
\pl{powerProp}(t, u, M)\ | \ \pl{powerRep}(t, u, M)
\]
\[
\pl{infProp}(t, M)\ | \ \pl{infRep}(t, M)
\]
\[
\pl{repl}_{\phi(a, b, \vec{f})}\pl{Prop}(t, u, \vec{u}, M)\ |\ \pl{repl}_{\phi(a,
b, \vec{f})}\pl{Rep}(t, u, \vec{u}, M)
\]
\[
\IND_{\phi(a, \vec{b})}(\vec{t}, M)
\]
The \pl{ind} term corresponds to the $\in$-induction axiom schema (IND${}_{\phi(a, \vec{f})}$),
and \pl{Prop} and \pl{Rep} terms correspond to the respective axioms.
The exact nature of the correspondence will become clear in the next
section. Briefly and informally, the \pl{Rep} terms are
\emph{representatives} of the fact that a $t$ is a member of a term
$t(\vec{u})$ and the \pl{Prop} terms provide the defining \emph{property} of
$t \in t(\vec{u})$. To avoid listing all of them every
time, we adopt a convention of using \pl{axRep} and \pl{axProp} terms to tacitly
mean all \pl{Rep} and \pl{Prop} terms, for \pl{ax} being one of \pl{empty}, \pl{pair}, \pl{union},
\pl{sep}, \pl{power}, \pl{inf} and \pl{repl}. With this convention in mind, we can summarize the
definition of the \pl{Prop} and \pl{Rep} terms as:
\[
\pl{axProp}(t, \vec{u}, M)\ |\ \pl{axRep}(t, \vec{u}, M), 
\]
where the number of terms in the sequence $\vec{u}$ depends on the particular
axiom. 

The free variables of a lambda term are defined as usual, taking into
account that variables in $\lambda$, \pl{case} and \pl{let} terms bind respective
terms. The relation of  $\alpha$-equivalence is defined taking this information into account. We consider $\alpha$-equivalent terms equal.
We denote the set of all free variables of a term $M$ by $FV(M)$ and the set
of the free first-order variables of a term by $FV_F(M)$. The free (first-order) variables of a context $\g$
are denoted by $FV(\g)$ ($FV_F(\g)$) and defined in a natural way. The
notation $M[x:=N]$ stands for a term $M$ with $N$ substituted for $x$. 
The set of all $\li$ lambda terms will be denoted by $\Lambda$. 

\subsection{Reduction rules}\label{rr}

The deterministic reduction relation $\to$ arises by lazily evaluating the
following base reduction rules:
\[
\Llap{(\lambda x : \phi.\ M)\ N \to M[x:=N]\quad}
\Rlap{\quad(\lambda a.\ M)\ t \to M[a:=t]}
\]
\[
\Llap{\FST(<M, N>) \to M \quad}
\Rlap{\quad\SND(<M, N>) \to N} 
\]
\[
\Llap{\CASE(\INL(M), x\!:\!\phi.\, N, x\!:\!\psi.\, O)\!\to\! N[x:=M]\quad}
\Rlap{\quad\CASE(\INR(M), x\!:\!\phi.\, N, x\!:\!\psi.\,O)\!\to\! O[x:=M]}
\]
\[
\LET\ [a, x : \phi] := [t, M]\ \IN\ N \to N[a:=t][x:=M]
\]
\[
\pl{axProp}(t, \vec{u}, \pl{axRep}(t, \vec{u}, M)) \to M 
\]
\[
\IND_{\phi(a, \vec{b})}(\vec{t}, M) \to \lambda c.\ M\ c\
(\lambda b. \lambda x : b \in c.\ \IND_{\phi(a, \vec{b})}(
\vec{t}, M)\ b)\mbox{\qquad $c, b, x$ new}
\]
The laziness is specified formally by the following evaluation contexts:
\[
[ \circ ] ::= \FST([ \circ ])\ |\ \SND([ \circ ])\ |\ \CASE([ \circ ], x :
\phi. M,
x : \psi.N)\ |\ \pl{axProp}(t, \vec{u}, [ \circ ])\ 
\]
\[
\LET\ [a, y : \phi] := [ \circ ]\ \IN\ N\ |\
[ \circ ]\ M\ |\ \pl{magic}([\circ])
\]
In other words, the (small-step) reduction relation arises from the base
reduction rules and the following inductive definition:
\[
\infer{\FST(M) \to \FST(M')}{M \to M'} \qquad
\infer{\SND(M) \to \SND(M')}{M \to M'}
\]
\[
\infer{\CASE(M, x : \phi.\ N, x : \psi.\ O) \to \CASE(M', x : \phi.\ N, x :
\psi.\ O)}{M \to M'}
\]
\[
\infer{\pl{axProp}(t, \vec{u}, M) \to \pl{axProp}(t, \vec{u}, M')}{M \to M'}
\qquad
\infer{\LET\ [a, y : \phi] := M\ \IN\ N \to \LET\ [a, y : \phi] := M'\ \IN\
N}{M \to M'}
\]
\[
\infer{M\ N \to M'\ N}{M \to M'} \qquad
\infer{\MAGIC(M) \to \MAGIC(M')}{M \to M'} 
\]

\begin{definition}
We write $M \downarrow$ if the reduction sequence starting from $M$
terminates. We write $M \downarrow v$ if we want to state that $v$
is the term at which this reduction sequence terminates. We write $M \to^* M'$ if
$M$ reduces to $M'$ in some number of steps. \todo{Better formulation?}
\end{definition}

We distinguish certain $\li$ terms as values. The values are generated
by the following abstract grammar, where $M$ is an arbitrary term. Clearly, there are
no reductions possible from values. 
\[
V ::= \lambda a.\ M\ |\ \lambda x : \phi.\ M\ |\ \INL(M)\ |\ \INR(M) \ |\
[t, M]\ |\ <M, N>\ |\ \pl{axRep}(t, \vec{u}, M)
\]

\subsection{Types}\label{lambdaa}

The type system for $\li$ is constructed according to the principle
of Curry-Howard isomorphism for \iizfr. Types are \izfr\ formulas.
Contexts, denoted by $\g$, are finite sets of pairs $(x_i, \phi_i)$, written
as $x_1 : \phi_1, {\ldots} , x_n : \phi_n$. The \emph{domain} of a context
$\g$ is the set $\{ x\ |\ (x, \phi) \in \g \}$ and it is denoted by $\DOM(\g)$.  The \emph{range} of a context $\g$ is the corresponding first-order logic context that contains
only formulas and is denoted by $rg(\g)$. The first group of rules
corresponds to the rules of IFOL:
\[
\infer{\g, x : \phi \p x : \phi}{} \qquad 
\infer[x \notin \DOM(\g)]{\gp \lambda x : \phi.\ M : \phi \to
\psi}{\g, x : \phi \p M : \psi} \qquad
\infer{\gp M\ N : \psi}{\gp M : \phi \to
\psi & \gp N : \phi} 
\]
\[
\infer{\gp <M, N> : \phi \land \psi}{\gp M : \phi & \gp N : \psi}
\qquad 
\infer{\gp \FST(M) : \phi}{\gp M : \phi \land \psi} \quad \infer{\gp \SND(M) :
\psi}{\gp M : \phi \land \psi} 
\]
\[
\infer{\gp \INL(M) : \phi \lor \psi}{\gp M : \phi} \qquad \infer{\gp \INR(M)
: \phi \lor \psi}{\gp M : \psi} 
\]
\[
\infer{\gp \CASE(M, x : \phi.\ N, x : \psi.\ O) : \vartheta}{\gp M :
\phi \lor \psi & \g, x : \phi \proves N : \vartheta & \g, x : \psi \proves O : \vartheta}
\]
\[
\infer[a \notin FV_F(\g)]{\gp \lambda a.\ M : \forall a.\ \phi}{\g \proves M : \phi}
\qquad 
\infer{\gp M\ t : \phi[a:=t]}{\gp M : \forall a.\ \phi}
\]
\[
\infer{\gp [t, M] : \exists a.\ \phi}{\gp M : \phi[a:=t]} \qquad 
\infer[a \notin FV_F(\g, \psi)]{\gp \LET\ [a, x : \phi] := M\ \IN\
N : \psi}{\gp M : \exists a.\ \phi & \g,  x : \phi \proves N : \psi} 
\]
\[
\infer{\gp \MAGIC(M) : \phi}{\gp M : \bot}
\]
The rest of the rules correspond to \iizfr\ axioms:
\[
\infer{\gp \pl{axRep}(t, \vec{u}, M) : t \in t_A(\vec{u})}{\gp M : \phi_A(t,
\vec{u})} \qquad \infer{\gp \pl{axProp}(t, \vec{u}, M) : \phi_A(t, \vec{u}) }{ \gp M : t \in
t_A(\vec{u})}
\]
\[
\infer{\gp \IND_{\phi(a, \vec{f})}(\vec{t}, M) : \forall a.\
\phi(a, \vec{t})}{\gp M : \forall c.\ (\forall b.\ b \in c \to \phi(b,
\vec{t})) \to \phi(c, \vec{t})} 
\]

\subsection{Properties of $\lz$}

We now prove a standard sequence of lemmas for $\la$. 

\begin{lemma}[Canonical Forms]
Suppose $M$ is a value and $\proves M : \vartheta$. Then: 
\begin{enumerate}[$\bullet$]
\item $\vartheta = t \in t_A(\vec{u})$ iff $M = \pl{axRep}(t, \vec{u}, N)$ and $\p N : \phi_A(t, \vec{u})$.
\item $\vartheta = \phi \lor \psi$ iff  ($M = \INL(N)$ and $\p N : \phi$) or ($M =
\INR(N)$ and $\p N : \psi$).
\item $\vartheta = \phi \land \psi$ iff $M = <N, O>$, $\p N : \phi$ and $\p
O : \psi$. 
\item $\vartheta = \phi \to \psi$ iff $M = \lambda x : \phi.\ N$ and $x :
\phi \p N : \psi$. 
\item $\vartheta = \forall a.\ \phi$ iff $M = \lambda a.\ N$ and $\p N :
\phi$.
\item $\vartheta = \exists a.\ \phi$ iff $M = [t, N]$ and $\p N : \phi[a:=t]$.
\item $\vartheta = \bot$ never happens.
\end{enumerate}
\end{lemma}
\proof
Immediate from the typing rules and the definition of values.\qed

\begin{lemma}[Weakening]
If $\gp M : \phi$ and $FV(\psi) \cup \{ x \}$ are fresh with respect to the proof tree $\gp M : \phi$, then $\g, x : \psi \p M : \phi$.
\end{lemma}
\proof
Straightforward induction on $\gp M : \phi$. The freshness assumption is
used in the treatment of the proof rules having side-conditions, such as 
introduction of the universal quantifier.\qed

There are two substitution lemmas, one for the propositional part, the other
for the first-order part of the calculus. Since the rules and terms of $\lz$
corresponding to \izfr\ axioms do not interact with substitutions in a
significant way, the proofs are routine. 

\begin{lemma}\label{lamsl}
If $\Gamma, x : \phi \proves M : \psi$ and  $\Gamma \proves N : \phi$, then
$\Gamma \proves M[x:=N] : \psi$.
\end{lemma}
\proof
By induction on $\g, x : \phi \p  M : \psi$. We show two interesting cases.
\begin{enumerate}[$\bullet$]
\item $\psi = \psi_1 \to \psi_2$, $M = \lambda y : \psi_1.\ O$. Using $\alpha$-conversion 
we can choose $y$ to be new, so that $y \notin FV(\g, x) \cup FV(N)$. The
proof tree must end with:
\[
\infer{\g, x : \phi \p \lambda y : \psi_1.\ O : \psi_1 \to \psi_2}{\g, x :
\phi, y : \psi_1 \p O : \psi_2}
\]
By the induction hypothesis, $\g, y : \psi_1 \p O[x:=N] : \psi_2$, so $\g \p \lambda y : \psi_1.\ O[x:=N] : \psi_1
\to \psi_2$. By the choice of $y$, $\gp (\lambda y : \psi_1.\ O)[x:=N] :
\psi_1 \to \psi_2$. 
\item $\psi = \psi_2, M = \LET\ [a, y : \psi_1] := M_1\ \IN\ M_2$. The proof tree ends with:
\[
\infer{\g, x : \phi \p \LET\ [a, y : \psi_1] := M_1\ \IN\
M_2 : \psi_2}{\g, x : \phi \p M_1 : \exists a.\ \psi_1 & \g, x : \phi, y : \psi_1 \p M_2
: \psi_2}
\]
Choose $a$ and $y$ to be fresh. By the induction hypothesis, $\gp M_1[x:=N] : \exists a.\ \psi_1$ and $\g, y
: \psi_1 \p M_2[x:=N] : \psi_2$. Thus $\gp \LET\ [a, y : \psi_1] :=
M_1[x:=N]\ \IN\ M_2[x:=N] : \psi_2$. By $a$ and $y$ fresh, $\gp (\LET\ [a, y : \psi_1] :=
M_1\ \IN\ M_2)[x:=N] : \psi_2$ which is what we want.\qed 
\end{enumerate}

\begin{lemma}\label{logsl}
If $\gp M : \phi$, then $\Gamma[a:=t] \proves M[a:=t] : \phi[a:=t]$.
\end{lemma}
\proof
By induction on $\gp M : \phi$. Most of the rules do not interact with
first-order substitution, so we show the proof just for the four of them which
do. 
\begin{enumerate}[$\bullet$]
\item $\phi = \forall b.\ \phi_1$, $M = \lambda b.\ M_1$. The proof tree ends with:
\[
\infer[b \notin FV_F(\g)]{\gp \lambda b.\ M_1 : \forall b.\ \phi_1}{\g \proves M_1 : \phi_1}
\]
Without loss of generality we can assume that $b \notin FV(t) \cup \{ a \}$. By the induction hypothesis, $\Gamma[a:=t] \proves M_1[a:=t] :
\phi_1[a:=t]$. Therefore $\g[a:=t] \p \lambda b.\ M_1[a:=t] : \forall
b.\ \phi_1[a:=t]$ and by the choice of $b$, $\g[a:=t] \p (\lambda b.\ M_1)[a:=t] \p (\forall b.\ \phi_1)[a:=t]$. 
\item $\phi = \phi_1[b:=u]$, $M = M_1\ u$ for some term $u$. The proof tree ends with:
\[
\infer{\gp M_1\ u : \phi_1[b:=u]}{\gp M_1 : \forall b.\ \phi_1}
\]
Choosing $b$ to be fresh, by the induction hypothesis we get $\g[a:=t] \p
M_1[a:=t] : \forall b.\ (\phi_1[a:=t])$, so $\g[a:=t] \p M_1[a:=t]\ u[a:=t] :
\phi_1[a:=t][b:=u[a:=t]]$. By Lemma \ref{formsubst} and $b \notin FV(t)$, we
get $\g[a:=t] \p (M_1\ u)[a:=t] : \phi_1[b:=u][a:=t]$.
\item 
\[
\infer{\gp [u, M] : \exists b.\ \phi}{\gp M : \phi[b:=u]}
\]
Choosing $b$ to be fresh, by the induction hypothesis we get $\g[a:=t] \p M[a:=t] : \phi[b:=u][a:=t]$. 
By Lemma \ref{formsubst} and $b \notin FV(t)$, we get $\g[a:=t] \p M[a:=t] :
\phi[a:=t][b:=u[a:=t]]$. Therefore $\g[a:=t] \p [u[a:=t], M[a:=t]] : \exists
b.\ \phi[a:=t]$, so also $\g[a:=t] \p ([u, M])[a:=t] : (\exists b.\
\phi)[a:=t]$.
\item
\[
\infer[b \notin FV_F(\g, \psi)]{\gp \LET\ [b, x : \phi] := M\ \IN\
N : \psi}{\gp M : \exists b.\ \phi & \g,  x : \phi \proves N : \psi} 
\]
We choose $b$ so that $b \notin FV(t)$. By the induction hypothesis $\g[a:=t] \p M[a:=t]
: \exists b.\ \phi[a:=t]$ and $\g[a:=t], x : \phi[a:=t] \p N[a:=t] :
\psi[a:=t]$. By our choice of $b$ and $b \notin FV_F(\g, \psi)$, we also
have $b \notin FV_F(\g[a:=t], \psi[a:=t])$. Thus also $\g[a:=t] \p \LET\  [b, x : \phi[a:=t]] := M[a:=t]\ \IN\ N[a:=t] :
\psi[a:=t]$.\qed
\end{enumerate}

With the lemmas at hand, Progress and Preservation easily follow:

\begin{lemma}[Subject Reduction, Preservation]
If $\gp M : \phi$ and $M \to N$, then $\gp N : \phi$.
\end{lemma}
\proof
By induction on the definition of $M \to N$. We show several cases. Case $M
\to N$ of:
\begin{enumerate}[$\bullet$]
\item $(\lambda x : \phi_1.\ M_1)\ M_2 \to M_1[x:=M_2]$. The term $M$ has
the form $M = (\lambda x : \phi_1.\ M_1)\ M_2$ and the proof 
proof tree $\gp M : \phi$ ends with:
\[
\infer{\gp (\lambda x : \phi_1.\ M_1)\ M_2 : \phi}
{
  \infer
  {
    \gp \lambda x : \phi_1.\ M_1 : \phi_1 \to \phi
  }
  {
    \g, x : \phi_1 \p M_1 : \phi
  }
  & 
  \gp M_2 : \phi_1
}
\]
By Lemma \ref{lamsl}, $\gp M_1[x:=M_2] : \phi_1$.
\item $\LET\ [a, x : \phi_1] := [t, M_1]\ \IN\ M_2 \to M_2[a:=t][x:=M_1]$.
The term $M$ has the form $M = \LET\ [a, x : \phi_1] := [t, M_1]\ \IN\ M_2$
and the proof tree $\gp M : \phi$ ends with:
\[
\infer
{\gp \LET\ [a, x : \phi_1] := [t, M_1]\ \IN\ M_2 : \phi}
{
  \infer{\gp [t, M_1] : \exists a.\ \phi_1}
  {\gp M_1 : \phi_1[a:=t]}
  & 
  \g, x : \phi_1 \proves M_2 : \phi
}
\]
Choose $a$ to be fresh. Thus $M_1[a:=t] = M_1$ and $\g[a:=t] = \g$. By the side-condition of the last
typing rule, $a \notin FV(\phi)$, so $\phi[a:=t] = \phi$. By Lemma
\ref{logsl} we get $\g[a:=t], x : \phi_1[a:=t] \p M_2[a:=t] : \phi[a:=t]$,
so also $\g, x : \phi_1[a:=t] \p M_2[a:=t] : \phi$. By Lemma \ref{lamsl}, we
get $\gp M_2[a:=t][x:=M_1] : \phi$.
\item $\pl{axProp}(t, \vec{u}, \pl{axRep}(t, \vec{u}, M_1)) \to M_1$. In this
case the term
$M$ is has the form  $M = \pl{axProp}(t, \vec{u}, \pl{axRep}(t, \vec{u},
M_1))$ and the proof tree ends with:
\[
\infer{\gp \pl{axProp}(t, \vec{u}, \pl{axRep}(t, \vec{u}, M_1)) : \phi_A(t, \vec{u})}
{
\infer{\gp \pl{axRep}(t, \vec{u}, M_1)) : t \in t_A(\vec{u})}{\gp M_1 :
\phi_A(t, \vec{u})}
}
\]
The claim follows immediately.
\item $\IND_{\psi(a, \vec{f})}(\vec{t}, M_1) \to \lambda c.\ M_1\ c\
(\lambda b. \lambda x : b \in c.\ \IND_{\psi(a, \vec{f})}(\vec{t},
M_1)\ b)$. The term $M$ has the form $M = \IND_{\psi(a,
\vec{f})}(\vec{t}, M_1)$ and the proof tree ends with:
\[
\infer{\gp \IND_{\psi(a, \vec{f})}(\vec{t}, M_1) : \forall a.\ \psi(a,
\vec{t})}{\gp M_1 : \forall c.\ (\forall b.\ b \in c \to \psi(b, \vec{t})) \to \psi(c, \vec{t})}
\]
We choose $b, c, x$ to be fresh. By applying $\alpha$-conversion we can also obtain a proof
tree of $\gp M_1 : \forall e.\ (\forall d.\ d \in e \to \psi(d, \vec{t}))
\to \psi(e, \vec{t})$, where $\{ d, e \} \cap \{ b, c \} = \emptyset$. Then
by Weakening we get $\g, x : b \in c \p M_1 : \forall e.\ (\forall d.\ d
\in e \to \psi(d, \vec{t})) \to \psi(e, \vec{t})$, so also $\g, x : b \in c \p
\IND_{\psi(a, \vec{f})}(\vec{t}, M_1)
: \forall a.\ \psi(a, \vec{t})$. Let the proof tree $T$ be defined as:
\[
\infer{\gp \lambda b. \lambda x : b \in c.\ \IND_{\psi(a,
\vec{f})}(\vec{t}, M_1)\ b : \forall b.\ b \in c \to \psi(b,
\vec{t})}
    {
      \infer{\gp \lambda x : b \in c.\ \IND_{\psi(a,
\vec{f})}(\vec{t}, M_1)\ b : b \in c \to \psi(b,
\vec{t})}
      {
        \infer{\g, x : b \in c \p \IND_{\psi(a, \vec{f})}(
\vec{t}, M_1)\ b : \psi(b, \vec{t})}
        {
	  \g, x : b \in c \p \IND_{\psi(a, \vec{f})}(\vec{t}, M_1)
: \forall a.\ \psi(a, \vec{t})
	}
      }
    }
\]
Then the following proof tree shows the claim:
\[
\infer{\gp \lambda c.\ M_1\ c\ (\lambda b. \lambda x : b \in c.\ \IND_{\psi(a,
\vec{f})}(\vec{t}, M_1)\ b) : \forall c.\ \psi(c, \vec{t})}
{
  \infer
  {
    \gp M_1\ c\ (\lambda b. \lambda x : b \in c.\ \IND_{\psi(a,
\vec{f})}(\vec{t}, M_1)\ b) : \psi(c, \vec{t})
  }
  {
    \infer{\gp M_1\ c : (\forall b.\ b \in c \to \psi(b, \vec{t})) \to \psi(c, \vec{t})}
    {
      \gp M_1 : \forall c.\ (\forall b.\ b \in c \to \psi(b, \vec{t})) \to
\psi(c, \vec{t})
    }
    &
    T
  }
}
\]\qed
\end{enumerate}

\begin{lemma}[Progress]
If $\ \proves M : \phi$, then either $M$ is a value or there is $N$ such that $M \to N$.
\end{lemma}
\proof
Straightforward induction on the length of $M$. We show the cases for the
terms corresponding to \izfr\ axioms.
\begin{enumerate}[$\bullet$]
\item If $M = \pl{axRep}(t, \vec{u}, N)$, then $M$ is a value.
\item If $M = \pl{axProp}(t, \vec{u}, O)$, then we have the following proof tree:
\[
\infer{\p \pl{axProp}(t, \vec{u}, O) : \phi_A(t, \vec{u})}
{
\p O : t \in t_A(\vec{u})
}
\]
By the induction hypothesis, either $O$ is a value or there is $O_1$ such
that $O \to O_1$. In the former case, by Canonical Forms, $O = \pl{axRep}(t,
\vec{u}, P)$ and $M \to P$. In the latter, by the evaluation rules $\pl{axProp}(t,
\vec{u}, O) \to \pl{axProp}(t, \vec{u}, O_1)$.
\item The $\pl{ind}$ terms always reduce.\qed
\end{enumerate}

\begin{corollary}\label{corlz}
If $\ \p M : \phi$ and $M \downarrow v$, then $\p v : \phi$ and $v$ is a value.
\end{corollary}

\begin{corollary}\label{corbot}
If $\p M : \bot$, then $M$ does not normalize.
\end{corollary}
\proof
If $M$ normalized, then by Corollary \ref{corlz} we would have a value of
type $\bot$, which by Canonical Forms is impossible.\qed

Finally, we state the formal correspondence between $\lz$ and \iizfr:

\begin{lemma}[Curry-Howard Isomorphism]\label{ch}
If $\gp  O : \phi$ then \iizfr $ + rg(\g) \p  \phi$, where $rg(\g) = \{
\phi\ |\ (x, \phi) \in \g \}$. If \iizfr $+ \g \p \phi$, then there exists a term $M$ such that $\og \p M :
\phi$, where $\og = \{ (x_\phi, \phi)\ |\ \phi \in \g \}$.
\end{lemma}
\proof
Both parts follow by easy induction on the proof. The first part is
straightforward, to get the claim simply erase the lambda terms from the
proof tree. For the second part, we show terms and trees corresponding to \iizfr\ axioms:
\begin{enumerate}[$\bullet$]
\item Let $\phi$ be one of the \iizfr\ axioms apart from $\in$-Induction.
Then $\phi = \forall \vec{a}.\ \forall c.\ c \in t_A(\vec{a}) \iffl \phi_A(c,
\vec{a})$ for the axiom (A). Recall that $\phi_1 \iffl \phi_2$ is an
abbreviation for $(\phi_1 \to \phi_2) \land (\phi_2 \to \phi_1)$. Let $M = \lambda x : c \in t_A(\vec{a}).\ \pl{axProp}(c, \vec{a}, x)$
and let $N = \lambda x : \phi_A(c, \vec{a}).\ \pl{axRep}(c, \vec{a}, x)$. 
Let $S$ be the following proof tree:
\[
    \infer
    {
    \gp M : c \in t_A(\vec{a}) \to \phi_A(c, \vec{a})
    }
    {
      \infer{\g, x : c \in t_A(\vec{a}) \p \pl{axProp}(c, \vec{a}, x) : \phi_A(c, \vec{a})}
      {
       \g, x : c \in t_A(\vec{a}) \p x : c \in t_A(\vec{a})
      }
    }
\]
And let $T$ be the following proof tree:
\[
    \infer{\gp N : \phi_A(c, \vec{a}) \to c \in t_A(\vec{a})}
    {
      \infer{\g, x : \phi_A(c, \vec{a}) \p \pl{axRep}(c, \vec{a}, x) : c \in t_A(\vec{a})}
      {
        \g, x : \phi_A(c, \vec{a}) \p x : \phi_A(c, \vec{a})
      }
    }
\]
Then the following proof tree shows the claim:
\[
\infer{\gp \lambda \vec{a} \lambda c. 
<M, N> : \forall \vec{a}.\ \forall c.\ c \in t_A(\vec{a}) \iffl \phi_A(c, \vec{a})}
{
  \infer
  {\gp <M, N> : c \in t_A(\vec{a}) \iffl \phi_A(c, \vec{a})}
  {
  S & T
  }
}
\]
\item Let $\phi$ be the $\in$-induction axiom. Let $M = \lambda \vec{f} \lambda x : (\forall a.
(\forall b.\ b \in a \to \psi(b, \vec{f})) \to \psi(a, \vec{f})).\
\IND_{\psi(a, \vec{f})}(\vec{f}, x)$.
The following proof tree shows the claim:
\[
\infer{\gp M : \forall \vec{f}. (\forall a. (\forall b.\ b \in a \to
\psi(b, \vec{f})) \to \psi(a, \vec{f})) \to \forall a.\ \psi(a, \vec{f})}
{
  \infer{\g, x : \forall a. (\forall b.\ b \in a \to \phi(b, \vec{f})) \to
\psi(a, \vec{f}) \p \IND_{\psi(a, \vec{f})}(\vec{f}, x) : \forall a.\ \psi(a, \vec{f})}
  {
    \g, x : \forall a. (\forall b.\ b \in a \to \psi(b, \vec{f})) \to
\psi(a, \vec{f}) \p x : \forall a. (\forall b.\ b \in a \to \psi(b, \vec{f})) \to
\psi(a, \vec{f})
  }
}
\]\qed
\end{enumerate}

Note that all proofs in this section are constructive and quite weak from
the proof-theoretic point of view --- Heyting Arithmetic should be
sufficient to formalize the arguments. However, by the Curry-Howard isomorphism
and Corollary \ref{corbot}, normalization of $\lz$ entails consistency of \iizfr,
which easily interprets Heyting Arithmetic. Therefore a normalization
proof must utilize much stronger means, which we introduce in the following
section. 

\newcommand\vl{V^{\lambda}}
\newcommand\vla{\vl_\alpha}
\newcommand\vlb{\vl_\beta}

\section{Realizability for \iizfr}\label{izfreal}

In this section we work in ZF. It is likely that \izfc\ would be sufficient, as excluded middle is not used explicitly; however, arguments using ordinals
and ranks would need to be done very carefully, as the notion of an ordinal
in constructive set theories is problematic \cite{powell, taylor96}.

Our definition of realizability is inspired by McCarty's presentation
in his Ph. D. thesis \cite{mccarty}. However, while he used it mainly to
prove independence results for \izfc\ and to carry out recursive mathematics,
we use it to prove normalization of $\lz$.

The realizability relation $\reals$ relates \emph{realizers} with \izfr\
formulas over an extended signature. The realizers are terms of $\lz$; the
signature is extended with class-many constants we call $\lambda$-names. We
proceed with the formal definitions.

\begin{definition}
The set of all values in $\lz$ is denoted by $\Lambda_{val}$. 
\end{definition}

\begin{definition}
A set $A$ is a $\lambda$-name iff $A$ is a set of pairs $(v, B)$ such that
$v \in \Lambda_{val}$ and $B$ is a $\lambda$-name.
\end{definition}

In other words, $\lambda$-names are sets hereditarily labelled by $\li$ values.

\begin{definition}
The class of $\lambda$-names is denoted by $\vl$.
\end{definition}

Formally, $V^\lambda$ is generated by the following transfinite inductive
definition on ordinals:
\[
V^\lambda_\alpha = \bigcup_{\beta \lt \alpha} P(\Lambda_{val} \times
V^\lambda_\beta) \qquad V^\lambda = \bigcup_{\alpha \in ORD}V^\lambda_\alpha
\]

The \emph{$\lambda$-rank} of a $\lambda$-name $A$, denoted by $\lrk(A)$, is the
smallest $\alpha$ such that $A \in \vla$.

\begin{definition}
For any $A \in \vl$, $A^+$ denotes $\{ (M, B)\ |\ M \downarrow v \land (v,
B) \in A \}$. 
\end{definition}

\begin{definition}
An \emph{environment} is a finite partial function from first-order
variables to $\vl$. 
\end{definition}
We will use the letter $\rho$ to denote \emph{environments}.

The environments are used to store elements of $\vl$. In order to smoothen the
presentation and make the account closer to the standard accounts of realizability for constructive set theories
\cite{mccarty,rathjendp,rathjenizf}, we make it possible for the formulas to mention constants 
from $\vl$ as well. Strictly speaking this is unnecessary and we could give
the account of the realizability relation and the normalization theorem using
only environments; the cost to pay would be some loss of clarity.

Formally, we extend the first-order language of \izfr\ in the following way:

\begin{definition}
A (class-sized) first-order language $L$ arises by enriching the \izfr\ signature
with constants for all $\lambda$-names.
\end{definition}

From now on until the end of this section, the letters $A, B, C$ range over $\lambda$-names. 

\begin{definition}
For any formula $\phi$ of $L$, any term $t$ of $L$ and $\rho$ defined on all free variables of
$\phi$ and $t$, we define by metalevel mutual induction a realizability relation $M
\reals_\rho \phi$ in an environment $\rho$ and a meaning of a term  
$\SB{t}_\rho$ in an environment $\rho$:
\begin{enumerate}[(1)]
\item $\SB{a}_\rho \equiv \rho(a)$
\item $\SB{A}_\rho \equiv A$
\item \label{omegadef} $\SB{\omega}_\rho \equiv \omega'$, where $\omega'$ is defined by the means
of inductive definition: $\omega'$ is the smallest set such that:
\begin{enumerate}[$\bullet$]
\item $(\pl{infRep}(\emptyset, N), A) \in \omega'$ if $N \downarrow \INL(O)$, $O \rrho A =
0$ and $A \in \vl_\omega$. 
\item If $(M, B) \in \omega'^+$, then $(\pl{infRep}(\emptyset, N), A) \in \omega'$ if $N
\downarrow \INR(N_1)$, $N_1 \downarrow [t, O]$, $O \downarrow <M, P>$, $P
\rrho A = S(B)$, $A \in \vl_\omega$. 
\end{enumerate}
Note that if $(M, B) \in \omega'^+$, then there is a finite ordinal $\alpha$
such that $B \in \vl_\alpha$.
\item \label{termdef} $\SB{t_A(\vec{u})}_\rho \equiv \{
(\pl{axRep}(\emptyset, \vec{\emptyset}, N),B) \in \Lambda_{val} \times
\vl_\gamma\ |\ N \reals_\rho \phi_A(B, \overrightarrow{\SB{u}_\rho})\}$
\item $M \reals_\rho \bot \equiv \bot$
\item $M \reals_\rho t \in s \equiv M \downarrow v \land (v, \SB{t}_\rho) \in \SB{s}_\rho$
\item $M \reals_\rho \phi \land \psi \equiv M \downarrow <M_1, M_2> \land M_1
\reals_\rho \phi \land M_2 \reals_\rho \psi$
\item $M \reals_\rho \phi \lor \psi \equiv (M \downarrow \INL(M_1) \land M_1
\reals_\rho \phi) \lor (M \downarrow \INR(M_1) \land M_1 \reals_\rho \psi)$
\item $M \reals_\rho \phi \to \psi \equiv (M \downarrow \lambda x.\ M_1) \land
\forall N.\ (N \reals_\rho \phi) \to (M_1[x:=N] \reals_\rho \psi)$
\item $M \reals_\rho \forall a.\ \phi \equiv M \downarrow \lambda a.\ N
\land \forall A \in V^\lambda, \forall t \in Tms.\ N[a:=t] \reals_\rho \phi[a:=A]$
\item $M \reals_\rho \exists a.\ \phi \equiv M \downarrow [t, N] \land \exists A \in
V^\lambda.\ N \reals_\rho \phi[a:=A]$
\end{enumerate}
\end{definition}

Note that $M \reals_\rho A \in B$ iff $(M, A) \in B^+$.

The definition of the ordinal $\gamma$ in item \ref{termdef} 
depends on $t_A(\vec{u})$. This ordinal is close to the rank of the set denoted
by $t_A(\vec{u})$ and is chosen so that Lemma \ref{realsterms} can be proven.
Let $\vec{\alpha} = \overrightarrow{\lrk(\SB{u}_\rho)}$.
Case $t_A(\vec{u})$ of:
\begin{enumerate}[$\bullet$]
\item $\emptyset$ --- $\gamma = \emptyset$. 
\item $\{ u_1, u_2 \}$ --- $\gamma = max(\alpha_1, \alpha_2)$. 
\item $P(u)$ --- $\gamma = \alpha + 1$.
\item $\bigcup u$ --- $\gamma = \alpha$. 
\item $S_{\phi(a, \vec{f})}(u, \vec{u})$ --- $\gamma = \alpha_1$.
\item $R_{\phi(a, b, \vec{f})}(u, \vec{u})$. This case is more complicated.
The names are chosen to match the corresponding clause in the proof of Lemma \ref{realsterms}. 
Let $G = \{ (N_1, (N_{21}, B)) \in \Lambda \times \SB{u}^+_\rho\ |\
\exists d \in \vl.\ \psi(N_1, N_{21}, B, d) \}$, where
$\psi(N_1, N_{21}, B, d) \equiv (N_1 \downarrow \lambda a.\ N_{11}) \land (N_{11}
\downarrow \lambda x.\ O) \land \exists s \in Tms.\ (O[x:=N_{21}] \downarrow [s, O_1]) \land (O_1 \reals_\rho
\phi(B, d, \overrightarrow{\SB{u}_\rho}) \land \forall e.\ \phi(B, e, \overrightarrow{\SB{u}_\rho}) \to
e = d)$. Then for all $g \in G$ there is $D$ and $(N_1, (N_{21}, B))$ such that $g =
(N_1, (N_{21}, B))$ and $\psi(N_1, N_{21}, B, D)$. Use Collection to collect these $D$'s in one set $H$, so that for
all $g \in G$ there is $D \in H$ such that the property holds. Apply Replacement
to $H$ to get the set of $\lambda$-ranks of sets in $H$. Then $\beta \equiv \bigcup H$ is
an ordinal and for any $D \in H$, $\lrk(D) \lt \beta$. Therefore for all $g \in G$ there is $D \in \vl_\beta$ and $(N_1, (N_{21}, B))$ such that $g =
(N_1, (N_{21}, B))$ and $\psi(N_1, N_{21}, B, D)$ holds. Set $\gamma = \beta + 1$.
\end{enumerate}

\begin{lemma}
The definition of realizability is well-founded. 
\end{lemma}
\proof
We define a measure function $m$ which takes a clause in the
definition and returns a triple of natural numbers:
\begin{enumerate}[$\bullet$]
\item $m(M \reals_\rho \phi)$ = (``number of constants $\omega$ in $\phi$'',
``number of function symbols in $\phi$'', ``structural complexity of $\phi$'')
\item $m(\SB{t}_\rho)$ = (``number of constants $\omega$ in $t$'', ``number of function symbols in $t$'', 0)
\end{enumerate}
With lexicographical order in $\nat^3$, it is trivial to check that the measure
of the definiendum is always greater than the measure of the definiens ---
the number of terms does not increase in the clauses for realizability and
the formula complexity goes down, in the clause for $\omega$, $\omega$ disappears and in the rest of clauses for terms,
the topmost $t_A$ disappears. Since $\nat^3$ with lexicographical order is
well-founded, the claim follows.\qed

Since the definition is well-founded, (metalevel) inductive proofs on the
definition of realizability are justified, such as the proof of the following lemma:

\begin{lemma}\label{realsubst}
$\SB{t[a:=s]}_\rho = \SB{t[a:=\SB{s}_\rho]}_\rho = \SB{t}_{\rho[a:=\SB{s}_\rho]}$ and $M \reals_\rho
\phi[a:=s]$ iff $M \reals_\rho \phi[a:=\SB{s}_\rho]$ iff $M \reals_{\rho[a:=\SB{s}_\rho]} \phi$.
\end{lemma}
\proof
Straightforward induction on the definition of realizability. We show representative
cases. Case $t$ of:
\begin{enumerate}[$\bullet$]
\item $A$ --- then $\SB{t[a:=s]}_\rho = \SB{t[a:=\SB{s}_\rho]}_\rho =
\SB{t}_{\rho[a:=\SB{s}_\rho]} = A$. 
\item $a$ --- then $\sr{t[a:=s]} = \sr{s}$, $\sr{t[a:=\sr{s}]} = \sr{\sr{s}}
= \sr{s}$ and also $\SB{t}_{\rho[a:=\sr{s}]} = \sr{s}$.
\item $t_A(\vec{u})$. Then $\sr{t[a:=s]} = \{ (\pl{axRep}(\emptyset,
\vec{\emptyset}, N), A)\
|\ N \rrho \phi_A(A, \vec{u}[a:=s]) \}$. By the induction
hypothesis, this set is equal to $\{ (\pl{axRep}(\emptyset,
\vec{\emptyset}, N), A)\ |\ N \rrho \phi_A(A, \vec{u}[a:=\sr{s}]) \} = \sr{t[a:=\sr{s}]}$
and also to 
$\{ (\pl{axRep}(\emptyset, \vec{\emptyset}, N), A)\ |\ 
N \reals_{\rho[a:=\sr{s}]} \phi_A(A, \vec{u}) \}$ and thus to 
$\SB{t}_{\rho[a:=\sr{s}]}$. 
\end{enumerate}
Case $\phi$ of:
\begin{enumerate}[$\bullet$]
\item $t \in u$. We have $M \rrho (t \in u)[a:=s]$ iff $M \rrho t[a:=s] \in
u[a:=s]$ iff $M \downarrow v$ and $(v, \sr{t[a:=s]}) \in \sr{u[a:=s]}$. By
the induction hypothesis, this is equivalent to $(v,
\SB{t[a:=\SB{s}_\rho]}_\rho) \in \SB{u[a:=\SB{s}_\rho]}_\rho$ and to
$(v, \SB{t}_{\rho[a:=\SB{s}_\rho]}) \in \SB{u}_{\rho[a:=\SB{s}_\rho]}$, so
also to $M \rrho t[a:=\SB{s}_\rho] \in
u[a:=\SB{s}_\rho]$ and to $M \reals_{\rho[a:=\SB{s}_\rho]} t
\in u$. This shows the claim.
\item $\forall b.\ \phi$. We have $M \rrho (\forall b.\ \phi)[a:=s]$ iff
(choosing $b$ to be fresh) $M \rrho \forall b.\ \phi[a:=s]$ iff $M
\downarrow \lambda b.\ N$ and $\forall A \in \vl, \forall u \in Tms.\ N[b:=u]
\rrho \phi[a:=s][b:=A]$. By the choice of $b$, this is equivalent to
$\forall A \in \vl, \forall u \in Tms.\ N[b:=u] \rrho \phi[b:=A][a:=s]$. By the induction hypothesis, this is equivalent to
$\forall A \in \vl, \forall u \in Tms.\ N[b:=u] \rrho \phi[b:=A][a:=\sr{s}]$
and to $\forall A \in \vl, \forall u \in Tms.\ N[b:=u] \reals_{\rho[a:=\sr{s}]}
\phi[b:=A]$, from which we easily recover the claim. \qed
\end{enumerate}

\begin{lemma}\label{realnorm}
If $(M \rrho \phi)$ then $M \downarrow$.
\end{lemma}
\proof
Straightforward from the definition of realizability. For $\phi = \bot$,
the claim trivially follows and in every other case the definition starts with a clause assuring normalization of $M$.\qed

\begin{lemma}\label{realredclosed}
If $M \to^* M'$ then $M'\reals_\rho \phi$ iff $M \reals_\rho \phi$.
\end{lemma}
\proof
Whether $M \rrho \phi$ or not depends only on the value of $M$, which does not
change with reduction or expansion.\qed

\begin{lemma}\label{afvreal}
If $\rho$ agrees with $\rho'$ on $FV(\phi)$, then $M \rrho \phi$ iff $M
\reals_{\rho'} \phi$. In particular, if $a \notin FV(\phi)$, then $M \rrho
\phi$ iff $M \reals_{\rho[a:=A]} \phi$. 
\end{lemma}
\proof
Straightforward induction on the definition of realizability --- the environment
is used only to provide the meaning of the free variables of terms in a
formula.\qed

\begin{lemma}\label{realimpl}
If $M \reals_\rho \phi \to \psi$ and $N \reals_\rho \phi$, then $M\ N \reals
\psi$. 
\end{lemma}
\proof
Suppose $M \reals_\rho \phi \to \psi$. Then $M \downarrow (\lambda x.\ O)$
and for all $P \reals \phi$, $O[x:=P] \reals \psi$. Now, $M\ N \to^*
(\lambda x.\ O)\ N \to O[x:=N]$. Lemma \ref{realredclosed} gives us the claim.\qed

We now prove a sequence of lemmas which culminates in Lemma
\ref{realsterms}, the keystone in the normalization proof. 

\begin{lemma}\label{ineqrank}
If $A \in \vla$ then there is $\beta \lt \alpha$ such that
for all $B$, if $M \reals_\rho B \in A$, then $B \in \vlb$. Also, if $M
\reals_\rho B = A$, then $B \in \vla$.
\end{lemma}
\proof
Take $A \in \vla$. Then there is $\beta \lt \alpha$ such that $A \in P(\Lambda_{val} \times
\vlb)$. Take any $B$. If $M \rrho B \in A$, then $M \downarrow v$ and $(v, B) \in A$, so $B \in \vlb$.

For the second part, suppose $M \reals_\rho A = B$. 
This means that $M \reals_\rho \forall c.\ c \in A \iffl c \in B$, so $M
\downarrow \lambda c.\ N$ and for all $t \in Tms$, for all $C$, $N[c:=t]
\rrho C \in A \iffl C \in B$, so $\forall t, C.\ N[c:=t] \downarrow
<M_1, M_2>$, $M_1 \rrho C \in A \to C \in B$ and $M_2
\rrho C \in B \to C \in A$. Thus, for all $t, C, M_2 \downarrow \lambda x.\ M_3$ and for
all $M_4 \rrho C \in B$, $M_3[x:=M_4] \rrho C \in A$. Take any element $(v, C) \in B$. Then $v
\rrho C \in B$, so $M_3[x:=v] \rrho C \in A$. Thus by the first part, $C \in \vlb$. 
Therefore $B \subseteq \Lambda_{val} \times \vlb$, so $B \in P(\Lambda_{val} \times \vlb) =
\vl_{\beta + 1}$, so $B \in \vla$.\qed

The following two lemmas will be used for the treatment of $\omega$ in Lemma
\ref{realsterms}.

\begin{lemma}\label{realunorderedpair}
If $A, B \in \vla$, then $\sr{\{ A, B \}} \in \vl_{\alpha + 1}$.
\end{lemma}
\proof
Take any $(M, C) \in \sr{\{ A, B \}}$. By the definition of $\sr{\{ A, B
\}}$, any such $C$ is in $\vla$, so $\sr{ \{ A, B \}} \in \vl_{\alpha + 1}$.\qed

\begin{lemma}\label{union}
If $A \in \vla$ and $(M, C) \in \sr{\bigcup A}$, then $C \in \vla$.
\end{lemma}
\proof
By the definition of $\sr{\bigcup A}$, if $(M, C) \in \sr{\bigcup A}$ then
$(M, C) \in \vl_{\lrk(A)}$, so $C \in \vla$.\qed

\begin{lemma}\label{realsucc}
If $A \in \vla$ and $M \rrho B = S(A)$, then $B \in \vl_{\alpha + 3}$.
\end{lemma}
\proof
$M \rrho B = S(A)$ means $M \rrho B = \bigcup \{ A, \{ A , A \} \}$.
By Lemma \ref{ineqrank}, it suffices to show that $\sr{\bigcup \{ A, \{ A ,
A \} \}} \in \vl_{\alpha + 3}$. Applying Lemma \ref{realunorderedpair}
twice, we find that $\sr{ \{ A, \{ A , A \} \}} \in \vl_{\alpha + 2}$. By
Lemma \ref{union}, if $(M, C) \in \sr{\bigcup \{ A, \{ A , A \} \}}$, then $C \in \vl_{\alpha + 2}$, which
shows the claim.\qed

The following lemma states the crucial property of the realizability relation.
\begin{lemma}\label{realsterms}
$(M, A) \in \SB{t_A(\vec{u})}_\rho$ iff $M = \pl{axRep}(\emptyset, \vec{\emptyset}, N)$ and $N \reals_\rho \phi_A(A,
\overrightarrow{\SB{u}_\rho})$. 
\end{lemma}
\proof
For all terms apart from $\omega$, the left-to-right part is immediate. For the
right-to-left part, suppose $N \reals_\rho \phi_A(A, \SB{\ov{u}}_\rho)$ and $M =
\pl{axRep}(\emptyset, \vec{\emptyset}, N)$. To show that $(M, A) \in \SB{t_A(\vec{u})}_\rho$, we need to show that $A
\in \vl_\gamma$.  The proof proceeds by case analysis
on $t_A(\vec{u})$. Let $\vec{\alpha} = \overrightarrow{\lrk(\SB{u}_\rho)}$. Case $t_A(\vec{u})$ of:
\begin{enumerate}[$\bullet$]
\item $\emptyset$. If $N \rrho \bot$ then anything holds, in particular $A \in \emptyset$. 
\item $\{ u_1, u_2 \}$. Suppose that $N \reals_\rho A = \SB{u_1}_\rho \lor A
= \SB{u_2}_\rho$. Then either $N \downarrow\ \INL(N_1) \land N_1 \reals_\rho A =
\SB{u_1}_\rho$ or $N \downarrow\ \INR(N_1) \land N_1 \reals_\rho A =
\SB{u_2}_\rho$. By Lemma \ref{ineqrank}, in the former case $A \in
\vl_{\alpha_1}$, in the latter $A \in \vl_{\alpha_2}$, so $A \in
\vl_{max(\alpha_1, \alpha_2)}$. 
\item $P(u)$. Suppose that $N \reals_\rho\forall c.\ c\ \in A \to c \in
\SB{u}_\rho$. Then $N \downarrow \lambda c.\ N_1$ and for all $t, C$,
$N_1[c:=t] \downarrow \lambda x.\ N_2$ and $\forall O.\ (O \reals
C \in A) \Rightarrow N_2[x:=O] \reals_\rho C \in \SB{u}_\rho$.
Take any $(v, B) \in A$. Then $v \reals_\rho B \in A$. So $N_2[x:=v]
\reals_\rho B \in \SB{u}_\rho$.
By Lemma \ref{ineqrank} any such $B$ is in $\vl_{\alpha}$, so $A \in \vl_{\alpha + 1}$.
\item $\bigcup u$. Suppose $N \reals_\rho\exists c.\ c \in \SB{u}_\rho \land A
\in c$. Then $N \downarrow [t, O]$ and there is $C$ such that $O \downarrow
<O_1, O_2>$, $O_1 \rrho C \in \sr{u}$ and $O_2 \rrho A \in C$. Two
applications of Lemma \ref{ineqrank} provide the claim. 
\item $S_{\phi(a, \vec{f})}(u, \vec{u})$. Suppose $N \reals_\rho A \in \SB{u}_\rho \land
\phi(A, \overrightarrow{\sr{u}})$. Then $N \downarrow <N_1, N_2>$ and $N_1 \rrho A \in
\sr{u}$. Lemma \ref{ineqrank} shows the claim. 
\item $R_{\phi(a, b, \vec{f})}(u, \vec{u})$. Suppose $N \reals_\rho (\forall x \in \SB{u}_\rho \exists! y.\ \phi(x,
y, \overrightarrow{\SB{u}_\rho})) \land \exists x \in \SB{u}_\rho.\ \phi(x, A, \overrightarrow{\SB{u}_\rho})$. Then $N \downarrow
<N_1, N_2>$ and $N_2 \reals_\rho\exists x \in \SB{u}_\rho.\ \phi(x, A,
\overrightarrow{\SB{u}_\rho})$. Thus $N_2 \downarrow [t, N_{20}]$, $N_{20} \downarrow
<N_{21}, N_{22}>$ and there is $B$ such that $N_{21} \reals_\rho B \in
\SB{u}_\rho$ and $N_{22} \reals_\rho\phi(B, A, \overrightarrow{\SB{u}_\rho})$. We also
have $N_1 \reals_\rho\forall x \in \SB{u}_\rho \exists! y.\ \phi(x, y,
\overrightarrow{\SB{u}_\rho})$, so $N_1 \downarrow \lambda a.\ N_{11}$ and for all $C,
t$, $N_{11}[a:=t] \downarrow
\lambda x.\ O$ and for all $P \rrho C \in \SB{u}_\rho$, $O[x:=P]
\rrho \exists !y.\ \phi(C, y, \overrightarrow{\SB{u}_\rho})$. So taking $C = B$, $t = a$
and $P=N_{21}$, there is $D$ such that $N_1 \downarrow \lambda a.\ N_{11}$,
$N_{11} \downarrow \lambda x.\ O$, $O[x:=N_{21}] \downarrow [s, O_1]$ and
$O_1 \rrho \phi(B, D, \overrightarrow{\SB{u}_\rho}) \land \forall e.\ \phi(B, e, \overrightarrow{\SB{u}_\rho}) \to e =
D$. Therefore $(N_1, (N_{21}, B)) \in G$ from the definition of $\gamma$, so 
there is $D \in V^{\lambda}_\gamma$ such that $N_1 \downarrow \lambda a.\
N_{11}$, $N_{11} \downarrow \lambda x.O$, $O[x:=N_{21}] \downarrow [s, O_1]$ and $O_1 \reals_\rho \phi(B,
D, \overrightarrow{\SB{u}_\rho}) \land \forall e.\ \phi(B, e, \overrightarrow{\SB{u}_\rho}) \to e =
D$. So $O_1 \downarrow <O_{11}, O_{12}>$ and $O_{12} \reals_\rho\forall e.\
\phi(B, e, \overrightarrow{\SB{u}_\rho}) \to e = D$. Therefore, $O_{12} \downarrow
\lambda a.\ Q$, $Q \downarrow \lambda x.\ Q_1$ (since we can take again $t = a$ and $Q[a:=a] = Q$) and $Q_1[x:=N_{22}]
\reals_\rho A = D$. By Lemma \ref{ineqrank}, $A \in \vl_\gamma$.
\end{enumerate}

Now we tackle $\omega$. For the left-to-right direction, obviously $M =
\pl{infRep}(\emptyset, N)$. For the claim about $N$, we proceed by induction on the
definition of $\omega'$:
\begin{enumerate}[$\bullet$]
\item The base case. Then $N \downarrow \INL(O)$ and $O \reals_\rho A = 0$, so $N
\reals_\rho A = 0 \lor \exists y \in \omega'.\ A = S(y)$. 
\item The inductive step. Then $N \downarrow \INR(N_1)$, $N_1 \downarrow [t, O]$,
$O \downarrow <M', P>$, $(M', B) \in \omega'^+$, $P \reals_\rho A = S(B)$.
Therefore, there is $C$ (namely $B$) such that $M' \reals_\rho C \in \omega'$ and $P
\reals_\rho A = S(C)$. Thus $[t, O] \reals_\rho \exists y.\ y \in
\omega' \land A = S(y)$, so $N \reals_\rho A = 0 \lor \exists y \in \omega'.\ A = S(y)$. 
\end{enumerate}
For the right-to-left direction, suppose $N \reals_\rho A = 0 \lor (\exists y.\ y \in
\omega'\land A = S(y))$. Then either $N \downarrow
\INL(N_1)$ or $N \downarrow \INR(N_1)$. In the former case, $N_1 \reals_\rho A =
0$, so by Lemma \ref{ineqrank} $A \in \vl_\omega$. In the latter, $N_1 \reals_\rho\exists y.\ y \in
\omega' \land A = S(y)$. Thus $N_1 \downarrow [t, O]$ and there is $B$ such that $O
\reals_\rho B \in \omega' \land A = S(B)$. So $O
\downarrow <M', P>$, $(M', B) \in \omega'^+$ and $P \reals_\rho A =
S(B)$. This is exactly the inductive step of the
definition of $\omega'$, so it remains to show that $A \in
\vl_\omega$. Since $(M', B) \in \omega'^+$, there is a finite ordinal
$\alpha$ such that $B \in \vl_\alpha$. By Lemma \ref{realsucc}, $A \in
\vl_{\alpha + 3}$, so also $A \in \vl_\omega$ and we get the claim.\qed

\section{Normalization}\label{sectionnorm}

In this section, environments $\rho$ are finite partial functions mapping 
proof variables to terms of $\la$ and first-order variables to pairs $(t,
A)$, where $t \in Tms$ and $A \in \vl$. Therefore, $\rho : Var \cup FVar \to
\Lambda \cup (Tms \times \vl)$, where $Var$ denotes the set of proof variables
and $FVar$ denotes the set of first-order variables. For any $\rho$,
$\rho_T$ denotes the restriction of $\rho$ to the mapping from first-order
variables into terms: $\rho_T = \lambda a \in FVar.\ \pi_1(\rho(a))$.  
Note that any $\rho$ can be used as a realizability environment by considering
only the mapping of first-order variables to $\vl$. 

We first define a reduction-preserving forgetting map $M \to \ov{M}$ on the terms of
$\lz$. The map changes all first-order arguments of $\pl{axRep}$ and
$\pl{axProp}$ terms to $\emptyset$. It is induced inductively in a natural way by the cases:
\[
\ov{\pl{axRep}(t, \vec{u}, M)} = \pl{axRep}(\emptyset, \vec{\emptyset}, \ov{M})
\qquad
\ov{\pl{axProp}(t, \vec{u}, M)} = \pl{axProp}(\emptyset, \vec{\emptyset}, \ov{M})
\]
So for example, $\ov{\lambda a.\ M} = \lambda a.\ \ov{M}, \ov{[t, M]} = [t,
\ov{M}], \ov{<M, N>} = <\ov{M}, \ov{N}>$ and so on. The reduction-preserving
character of the map is captured by the following lemmas:

\begin{lemma}\label{en1}
If $M \to N$ then $\ov{M} \to \ov{N}$. 
\end{lemma}
\proof
Straightforward. The first-order terms mapped to $\emptyset$ do not play a
role in reductions.\qed

\begin{lemma}\label{erasurenorm}
If $\ov{M}$ normalizes, then so does $M$. 
\end{lemma}
\proof
By Lemma \ref{en1}, an infinite reduction sequence starting from $M$ would
induce an infinite reduction sequence starting from $\ov{M}$.\qed

\begin{definition}
For a sequent $\gp \phi$, $\rho \models \gp M : \phi$ means that $\rho$ is
defined on $FV(\Gamma, M, \phi)$ and for all $(x_i, \phi_i) \in \g$, $\rho(x_i) \reals_\rho \phi_i$.
\end{definition}

Note that if $\rho \models \gp M : \phi$, then for any term $t$ in $\g, \phi$,
$\SB{t}_{\rho}$ is defined and so is the realizability relation $M
\reals_{\rho} \phi$.

\begin{definition}
For a sequent $\gp M : \phi$, if $\rho \models \gp M : \phi$ then $M[\rho]$
is $M[x_1 := \rho(x_1), {\ldots} , x_n := \rho(x_n), a_1:=\rho_T(a_1),
{\ldots}, a_k:=\rho_T(a_k)]$, where $FV(M) = \{ x_1, {\ldots}, x_n \}$ and $FV_F(M) = \{ a_1, {\ldots} , a_k \}$.
Similarly, if $\rho$ is defined on the free variables $a_1, {\ldots} , a_k$
of $t$, then $t[\rho]$ denotes $t[a_1:=\rho_T(a_1), {\ldots} ,
a_k:=\rho_T(a_k)]$. 
\end{definition}

\begin{lemma}\label{rhosubst}
If $\rho$ is not defined on $x$, then $M[\rho][x:=N] = M[\rho[x:=N]]$. Also
if $\rho$ is not defined on $a$, then $M[a:=t] = M[\rho[a:=(t, A)]]$. 
\end{lemma}
\proof
Straightforward structural induction on $M$.\qed

\begin{thm}[Normalization]\label{norm}
If $\gp M : \vartheta$ then for all $\rho \models \gp M : \vartheta$, $\ov{M}[\rho] \reals_\rho \vartheta$.
\end{thm}
\proof
For any $\la$ term $M$, $M'$ in the proof denotes $\ov{M}[\rho]$.
We proceed by metalevel induction on $\gp M : \vartheta$. Case $\gp M : \vartheta$ of:
\begin{enumerate}[$\bullet$]
\item 
\[
\infer{\g, x : \phi \proves x : \phi}{}
\]
Then $M' = \rho(x)$ and the claim follows.
\item 
\[
\infer{\gp M\ N : \psi}{\gp M : \phi \to \psi & \gp N : \phi}
\]
By the induction hypothesis, $M' \reals_\rho \phi \to \psi$ and $N' \reals_\rho \phi$. Lemma
\ref{realimpl} gives the claim.
\item
\[
\infer{\gp \lambda x : \phi.\ M : \phi \to \psi}{\g, x : \phi \p M : \psi}
\]
Take any $\rho \models \g$ and fresh $x$. We need to show that for any $N \reals_\rho
\phi$, $M'[x:=N] \reals_\rho \psi$. Take any such $N$. Let
$\rho' = \rho[x:=N]$. Then $\rho' \models \Gamma, x : \phi \p M : \psi$, so by the
induction hypothesis $\ov{M}[\rho'] \reals_{\rho'} \psi$. Since $x$ is
fresh, $\rho$ is undefined on $x$, so by Lemma \ref{rhosubst} $\ov{M}[\rho']
= \ov{M}[\rho][x:=N] = M'[x:=N]$. Therefore $M'[x:=N] \reals_{\rho'}
\psi$. Since $\rho'$ agrees with $\rho$ on logic variables, by Lemma \ref{afvreal} we get $M'[x:=N] \reals_\rho \psi$.
\item 
\[
\infer{\gp \pl{magic}(M) : \phi}{\gp M : \bot}
\]
By the induction hypothesis, $M' \rrho \bot$, which is not the case, so
anything holds, in particular $\pl{magic}(M') \reals_\rho \phi$.
\item
\[
\infer{\gp \FST(M) : \phi}{\gp M : \phi \land \psi}
\]
By the induction hypothesis, $M' \reals_\rho \phi \land \psi$, so $M' \downarrow <M_1, M_2>$ and
$M_1 \reals_\rho \phi$. Therefore $\FST(M) \to^* \FST(<M_1, M_2>) \to M_1$.
Lemma \ref{realredclosed} gives the claim. 
\item 
\[
\infer{\gp \SND(M) : \psi}{\gp M : \phi \land \psi}
\]
Symmetric to the previous case. 
\item 
\[
\infer{\gp <M, N> : \phi \land \psi}{\gp M : \phi & \gp N : \psi}
\]
All we need to show is $M' \reals_\rho \phi$ and $N' \reals_\rho \psi$, which we
get from the induction hypothesis.
\item
\[
\infer{\gp \INL(M) : \phi \lor \psi}{\gp M : \phi}
\]
We need to show that $M' \reals_\rho \phi$, which we get from the induction hypothesis.
\item
\[
\infer{\gp \INR(M) : \phi \lor \psi}{\gp M : \psi}
\]
Symmetric to the previous case.
\item 
\[
\infer{\gp \CASE(M, x : \phi.\ N, x : \psi.\ O) : \vartheta}{\gp M :
\phi \lor \psi & \g, x : \phi \proves N : \vartheta & \g, x : \psi \proves O : \vartheta}
\]
By the induction hypothesis, $M' \rrho \phi \lor \psi$. Take $x$ fresh, so
that $\rho$ is undefined on $x$. Therefore either $M'
\downarrow \INL(M_1)$ and $M_1 \rrho \phi$ or $M' \downarrow \INR(M_2)$ and
$M_2 \rrho \psi$. We only treat the former case, the latter is symmetric.
Since $\rho[x:=M_1] \rrho \g, x : \phi \proves N : \vartheta$, by the
induction hypothesis we get $\ov{N}[\rho[x:=M_1]] \rrho \vartheta$. We also
have $\CASE(M, x : \phi.\ \ov{N}, x : \psi.\ \ov{O}) \to^* \CASE(\INL(M_1), x :
\phi.\ \ov{N}, x : \psi.\ \ov{O}) \to \ov{N}[x:=M_1]$. By Lemma
\ref{rhosubst}, $\ov{N}[x:=M_1] =
\ov{N}[\rho[x:=M_1]]$, so Lemma \ref{realredclosed} gives us the claim.
\item
\[
\infer{\gp \lambda a.\ M : \forall a.\ \phi}{\gp M : \phi}
\]
By the induction hypothesis, for all $\rho' \models \gp M : \phi$, $M[\rho']
\reals_{\rho'} \phi$. We need to show that for all $\rho \models \gp \lambda a.\ M :
\forall a.\ \phi$, $\ov{(\lambda a.\ M)}[\rho] \reals_\rho \forall a.\
\phi$. Take any such $\rho$.  Using $\alpha$-conversion we can assure that $\rho$ is not defined on
$a$, so it suffices to show that $\lambda a.\ \ov{M}[\rho] \rrho \forall a.\ \phi$,
which is equivalent to $\forall A, t.\ \ov{M}[\rho][a:=t] \reals_\rho
\phi[a:=A]$. Take any $A$ and $t$. By Lemma \ref{realsubst} it suffices to
show that $\ov{M}[\rho][a:=t] \reals_{\rho[a:=A]} \phi$. 
Since $\rho[a:=(t, A)] \models \gp M : \phi$, by the induction hypothesis we get 
$\ov{M}[\rho[a:=(t, A)]] \reals_{\rho[a:=A]} \phi$. By Lemma \ref{rhosubst}
$\ov{M}[\rho][a:=t] =
\ov{M}[\rho[a:=(t, A)]]$, which shows the claim. 
\item
\[
\infer{\gp M\ t : \phi[a:=t]}{\gp M : \forall a.\ \phi}
\]
By the induction hypothesis, $M' \reals_\rho \forall a.\ \phi$, so $M' \downarrow \lambda a.\ N$
and $\forall A, u.\ N[a:=u] \reals_\rho \phi[a:=A]$. In particular $N[a:=t[\rho]]
\rrho \phi[a:=\sr{t}]$. By Lemma \ref{realsubst}, $N[a:=t[\rho]] \reals_\rho
\phi[a:=t]$. Since $\ov{M\ t}[\rho] = M'\ (t[\rho]) \to^* (\lambda a.\ N)\ t[\rho] \to
N[a:=t[\rho]]$, Lemma \ref{realredclosed} gives us  the claim.
\item 
\[
\infer{\gp [t, M] : \exists a.\ \phi}{\gp M : \phi[a:=t]}
\]
By the induction hypothesis, $M' \reals_\rho \phi[a:=t]$, so by Lemma
\ref{realsubst}, $M' \reals_\rho \phi[a:=\SB{t}_\rho]$. Thus, there is a $\lambda$-name $A$, namely $\SB{t}_\rho$, such that $M' \reals_\rho \phi[a:=A]$. Thus,
$[t, M][\rho]=[t[\rho], M'] \reals_\rho \exists a.\ \phi$, which is what we want.
\item
\[
\infer[a \notin FV(\Gamma, \psi)]{\gp \LET\ [a, x : \phi] := M\ \IN\ N : \psi}
{\gp M : \exists a.\ \phi & \g, x : \phi \proves N : \psi}
\]
Let $\rho \models \gp \LET\ [a, x : \phi] := M\ \IN\ N : \psi$. Choose $x,
a$ so that $\rho$ is undefined on these variables. 
We need to
show $\ov{(\LET\ [a, x : \phi ] := M\ \IN\ N)}[\rho] = \LET\ [a, x : \phi] := M'\ \IN\ 
\ov{N}[\rho] \rrho \psi$.
By the induction hypothesis, $M' \rrho \exists a.\ \phi$, so $M' \downarrow [t, M_1]$ and
for some $A$, $M_1 \rrho \phi[a:=A]$. By the induction hypothesis again, for any $\rho' \models \g,
x : \phi \p N : \psi$ we have $\ov{N}[\rho'] \reals_{\rho'} \psi$. Take
$\rho' = \rho[x:=M_1, a:=(t, A)]$. Since $a \notin FV(\psi)$, by Lemma
\ref{afvreal} $\ov{N}[\rho'] \rrho \psi $. Now, $\LET\ [a, x : \phi] := M'\
\IN\ \ov{N}[\rho] \to^* \LET\ [a, x : \phi] :=
[t, M_1]\ \IN\ \ov{N}[\rho] \to \ov{N}[\rho][a:=t][x:=M_1] = \ov{N}[\rho']$.
Lemma \ref{realredclosed} gives us the claim.
\item 
\[
\infer{\gp \pl{axRep}(t, \vec{u}, M) : t \in t_A(\vec{u})}{\gp M : \phi_A(t, \vec{u})}
\]
By the induction hypothesis, $M' \reals_\rho \phi_A(t, \vec{u})$. By Lemma \ref{realsubst} 
this is equivalent to $M' \reals_\rho \phi_A(\SB{t}_\rho, \ov{\SB{u}_\rho})$.
By Lemma \ref{realsterms}, $(\pl{axRep}(\emptyset, \vec{\emptyset}, M'), \SB{t}_\rho) \in
\SB{t_A(\vec{u})}_\rho$, so $\ov{\pl{axRep}(t, \vec{u}, M)} \rrho t \in t_A(\vec{u})$. 
\item
\[
\infer{\gp \pl{axProp}(t, \vec{u}, M) : \phi_A(t, \vec{u}) }{ \gp M : t
\in t_A(\vec{u})}
\]
By the induction hypothesis, $M' \reals_\rho t \in t_A(\vec{u})$. This means that 
$M' \downarrow v$ and $(v, \SB{t}_\rho) \in \sr{t_A(\vec{u})}$. 
By Lemma \ref{realsterms}, $v = \pl{axRep}(\emptyset, \vec{\emptyset}, N)$ and $N \reals_\rho \phi_A(\SB{t}_\rho, \ov{\SB{u}_\rho})$.
By Lemma \ref{realsubst}, $N \reals_\rho \phi_A(t, \vec{u})$.
Moreover, 
\[\ov{\pl{axProp}(t, \vec{u}, M)}[\rho] = \pl{axProp}(\emptyset,
\vec{\emptyset}, M') \to^* \pl{axProp}(\emptyset, \vec{\emptyset}, \pl{axRep}(
\emptyset, \vec{\emptyset}, N)) \to N\ .\]
 Lemma \ref{realredclosed} gives us the claim.
\item
\[
\infer{\gp \IND_{\phi(a, \vec{f})}(\vec{t}, M) : \forall a.\
\phi(a, \vec{t})}{\gp M : \forall c.\ (\forall b.\ b \in c \to \phi(b,
\vec{t})) \to \phi(c, \vec{t})}
\]
Since $\IND_{\phi(a, \vec{f})}(\vec{t}, M')$ reduces to $\lambda c.\ M'\ c\ (\lambda b.\ \lambda x.\
\IND_{\phi(a, \vec{f})}(\vec{t}, M')\ b)$, by Lemma \ref{realredclosed} it suffices to show that for all $C, t$,
$M'\ t\ (\lambda b.\ \lambda x.\ \IND_{\phi(a, \vec{f})}(\vec{t}, M')\ b) \rrho \phi(C,
\vec{t})$. We proceed by induction on $\lambda$-rank of $C$. Take any $C, t$. 
By the induction hypothesis, $M' \reals_\rho \forall c.\ (\forall b.\ b \in c \to \phi(b, \vec{t}))
\to \phi(c, \vec{t})$, so $M' \downarrow \lambda c.\ N$ and $N[c:=t] \rrho (\forall
b.\ b \in C \to \phi(b, \vec{t})) \to \phi(C, \vec{t})$. By Lemma
\ref{realredclosed}, $M'\ t \rrho (\forall
b.\ b \in C \to \phi(b, \vec{t})) \to \phi(C, \vec{t})$, so 
by Lemma \ref{realimpl}, it suffices to
show that $\lambda b.\ \lambda x.\ \IND_{\phi(a, \vec{f})}(\vec{t}, M')\ b \rrho \forall b.\ b \in C \to \phi(b, \vec{t})$.
Take any $B, u$, $O \rrho B \in C$, we need to show that
$\IND_{\phi(a, \vec{f})}(\vec{t}, M')[x:=O]\ u \rrho \phi(B, \vec{t})$. As $x \notin FV(M')$, it suffices
to show that $\IND_{\phi(a, \vec{f})}(\vec{t}, M')\ u \rrho
\phi(B, \vec{t})$, which, by Lemma \ref{realredclosed}, is equivalent to $M'\
u\ (\lambda b.\ \lambda x.\ \IND_{\phi(a, \vec{f})}(\vec{t}, M')\ b) \rrho \phi(B, \vec{t})$. 
As $O \rrho B \in C$, the $\lambda$-rank of $B$ is less than the
$\lambda$-rank of $C$ and we get the claim by the induction hypothesis.\qed
\end{enumerate}

\begin{corollary}[Normalization]\label{cornorm}
If $\proves M : \phi$, then $M \downarrow$. 
\end{corollary}
\proof
Take $\rho$ mapping all free proof variables of $M$ to themselves
and all free first-order variables $a$ of $M$ to $(a, \emptyset)$. 
Then $\rho \models \p M : \phi$. By Theorem \ref{norm}, $\ov{M}[\rho]$
normalizes. By the definition of $\rho$, $\ov{M}[\rho] = \ov{M}$. By Lemma
\ref{erasurenorm}, $M$ normalizes.\qed

Recall that in non-deterministic reduction systems, strong normalization means
that for any term $M$, all reduction paths starting from $M$ terminate, while weak normalization means
that for any term $M$ there is a terminating reduction path starting from
$M$. Our reduction system for $\li$ can be viewed as selecting a call-by-need reduction
strategy in a non-deterministic reduction system, where a reduction can be
applied anywhere inside of the term. In this view, our results show only
weak normalization of the calculus. Strong normalization then, surprisingly, does not hold. One reason, trivial, are
$\IND$ terms. However, even without them, the system would not strongly
normalize, as the following counterexample, invented by M. Crabb\'e and adapted to our framework shows:

\begin{thm}[Crabb\'e's counterexample]
There is a formula $\phi$ and a term $M$ such that $\p M : \phi$ and $M$
does not strongly normalize.
\end{thm}
\proof
Let $t = \{ x \in 0\ |\ x \in x \to \bot \}$. Consider the terms:
\[
N \equiv \lambda y : t \in t.\ \SND(\pl{sepProp}(t, 0, y))\ y \qquad
M \equiv \lambda x : t \in 0.\ N\ (\pl{sepRep}(t, 0, <x, N>))
\]
We first show that these terms can be typed. Let $T$ denote the following proof tree, showing that 
$\p N : t \in t \to \bot$:
\[
\infer{\p \lambda y : t \in t.\ \SND(\pl{sepProp}(t, 0, y))\ y : t \in t \to \bot}
{
  \infer{y : t \in t \p \SND(\pl{sepProp}(t, 0, y))\ y : \bot}
  {
    \infer{y : t \in t \p \SND(\pl{sepProp}(t, 0, y)) : t \in t \to \bot}
    {
      \infer{y : t \in t \p \pl{sepProp}(t, 0, y)) : t \in 0 \land t \in t \to \bot}
      {
        \infer{y : t \in t \p y : t \in \{ x \in 0 \ | \ x \in x \to \bot \}}{}
      }
    }
    &
    \infer{y : t \in t \p y : t \in t}{}
  }
}
\]
By Weakening, we can also obtain a tree $T_1$ showing that $x : t \in
0 \p N : t \in t \to \bot$. The following proof tree shows that $\p M : t \in 0 \to \bot$:
\[
\infer{\p \lambda x : t \in 0.\ N\ (\pl{sepRep}(t, 0, <x, N>)) : t \in 0 \to \bot}
{
  \infer{x : t \in 0 \p N\ (\pl{sepRep}(t, 0, <x, N>)) : \bot}
  {
    \infer{x : t \in 0 \p N : t \in t \to \bot}{T_1}
    &
    \infer{x : t \in 0 \p \pl{sepRep}(t, 0, <x, N>) : t \in t}
    {
      \infer{x : t \in 0 \p <x, N> : t \in 0 \land t \in t \to \bot}
      {
        \infer{x : t \in 0 \p x : t \in 0}{}
        &
        \infer{x : t \in 0 \p N : t \in t \to \bot}{T_1}
      }
    }
  }
}
\]
We now exhibit an infinite reduction sequence starting from $M$:
\[
\begin{array}{ll}
M = \lambda x : t \in 0.\ N\ (\pl{sepRep}(t, 0, <x, N>)) & = \\
\lambda x : t \in 0.\ (\lambda y : t \in t.\ \SND(\pl{sepProp}(t, 0, y))\
y)\ (\pl{sepRep}(t, 0, <x, N>)) & \to\\
\lambda x : t \in 0.\ \SND(\pl{sepProp}(t, 0, (\pl{sepRep}(t, 0, <x, N>))))\
(\pl{sepRep}(t, 0, <x, N>)) & \to\\
\lambda x : t \in 0.\ \SND(<x, N>)\ (\pl{sepRep}(t, 0, <x, N>)) & \to \\
\lambda x : t \in 0.\ N\ (\pl{sepRep}(t, 0, <x, N>)) = M & \to {\ldots}
\end{array}
\]\qed
Note that the counterexample also shows that the weak
normalization of $\lz$ is really weak --- although $\p M : \phi$ entails weak 
normalization of $M$, $\gp M : \phi$ does not, as there is a context $\g$ such that 
$\gp M : \phi$ and $M$ does not normalize. 

Moreover, a slight (from a semantic point of view) modification to \iizfr,
namely making it non-well-founded, results in a system which is not even
weakly normalizing. A very small fragment is sufficient for this effect to
arise. Let $T$ be an intuitionistic set theory consisting of 2 axioms:

\begin{enumerate}[$\bullet$]
\item (C) $\forall a.\ a \in c \iffl a = c$
\item (D) $\forall a.\ a \in d \iffl a \in c \land a \in a \to a \in a$.
\end{enumerate}

The constant $c$ denotes a non-well-founded set. The existence of $d$ can
be derived from the Separation axiom: $d = \{ a \in c\ | \ a \in a \to a \in a
\}$. The lambda calculus corresponding to $T$ is defined just as for \iizfr.

\begin{lemma}\label{dc}
$T \p d \in c$
\end{lemma}
\proof
It suffices to show that $d = c$. Take any $e \in d$, then $e \in c$. On the
other hand, suppose $e \in c$. Since obviously $e \in e \to e \in e$, we
also get $e \in d$.\proof

\begin{thm}\label{notweakly}
There is a formula $\phi$ and a term $M$ such that $\p_T M : \phi$ and $M$
does not weakly normalize.
\end{thm}
\proof
Let $N$ be the lambda term corresponding to the proof of Lemma \ref{dc} along
with the proof tree $T_N$. Take $\phi = d \in d \to d \in d$. Consider the terms:
\[
O \equiv \lambda x : d \in d.\ \SND(\pl{dProp}(d, c, x))\ x
\qquad 
M \equiv O\ (\pl{dRep}(d, c, <N, O>)).
\]
Again, we first show that these terms are typable. Let $S$ be the following proof
tree, showing that $\p O : d \in d \to d \in d$:
\[
\infer{\p \lambda x : d \in d.\ \SND(\pl{dProp}(d, c, x))\ x : d \in d \to d
\in d}
{
  \infer{x : d \in d \p \SND(\pl{dProp}(d, c, x))\ x : d \in d}
  {
    \infer{x : d \in d \p \SND(\pl{dProp}(d, c, x)) : d \in d \to d \in d}
    {
      \infer{x : d \in d \p \pl{dProp}(d, c, x)) : d \in c \land d \in d \to
d \in d}
      {
        \infer{x : d \in d \p x : d \in d}{}
      }
    }
    &
    \infer{x : d \in d \p x : d \in d}{}
  }
}
\]
Then the following proof tree shows that $M$ is typable:
\[
\infer{\p O\ (\pl{dRep}(d, c, <N, O>)) : d \in d}
{
  \infer{\p O : d \in d \to d \in d}{S}
  &
  \infer{\p \pl{dRep}(d, c, <N, O>) : d \in d}
  {
    \infer{\p <N, O> : d \in c \land d \in d \to d \in d}
    {
      \infer{\p N : d \in c}{T_N}
      &
      \infer{\p O : d \in d \to d \in d}{S}
    }
  }
}
\]
Finally, we exhibit the only reduction sequence starting from $M$:
\[
\begin{array}{ll}
M = O\ (\pl{dRep}(d, c, <N, O>)) & = \\
(\lambda x : d \in d.\ \SND(\pl{dProp}(d, c, x))\ x)\ (\pl{dRep}(d, c, <N,
O>)) & \to\\
\SND(\pl{dProp}(d, c, \pl{dRep}(d, c, <N,
O>)))\ (\pl{dRep}(d, c, <N, O>)) & \to\\
\SND(<N, O>)\ (\pl{dRep}(d, c, <N, O>)) & \to\\
O\ (\pl{dRep}(d, c, <N, O>)) = M & \to {\ldots} 
\end{array}
\]\qed

These counterexamples to normalization properties can also be presented in a
cleaner way in the framework of higher-order rewriting \cite{jawst2006}. 

\section{Applications}\label{secapp}

The normalization theorem immediately provides several results. 

\begin{corollary}[Disjunction Property]
If \iizfr $\p \phi \lor \psi$, then \iizfr $\p \phi$ or \iizfr $\p \psi$. 
\end{corollary}
\proof
Suppose \iizfr $\p \phi \lor \psi$. By the Curry-Howard isomorphism, there is a
$\li$ term $M$ such that $\p M : \phi \lor \psi$. By Corollary
\ref{corlz}, $M \downarrow v$ and $\p v : \phi \lor \psi$. By
Canonical Forms, either $v = \INL(N)$ and $\p N : \phi$ or $v = \INR(N)$
and $\p N : \psi$. By applying the other direction of the Curry-Howard isomorphism
we get the claim.\qed

\begin{corollary}[Term Existence Property]
If \iizfr $\p \exists x.\ \phi(x)$, then there is a closed term $t$ such that \iizfr $\p
\phi(t)$. 
\end{corollary}
\proof
By the Curry-Howard isomorphism, there is a $\li$-term $M$ such that $\p M :
\exists x.\ \phi$. By normalizing $M$ and applying
Canonical Forms, we get $[t, N]$ such that $\p N : \phi(t)$ and thus by
the Curry-Howard isomorphism \iizfr $\p \phi(t)$. If $t$ is not closed already, 
then let $\vec{a} = FV(t)$. We have \iizfr $\p \forall \vec{a}.\ \phi(t)$,
so also $\phi(t[\vec{a}:=\vec{\emptyset}])$.\qed

To show NEP, we first define an extraction function $F$ 
which takes a proof $\p M : t \in \omega$ and returns a natural number $n$.
$F$ works as follows:

It normalizes $M$ to $\pl{natRep(t, N)}$. By Canonical Forms, $\p N
: t = 0 \lor \exists y \in \omega.\ t = S(y)$. $F$ then normalizes $N$ to
either $\INL(O)$ or $\INR(O)$. In the former case, $F$ returns $0$. In the
latter, $\p O : \exists y.\ y \in \omega \land t = S(y)$. Normalizing $O$ it
gets $[t_1, P]$, where $\p P : t_1 \in \omega \land t = S(t_1)$. Normalizing
$P$ it obtains $Q$ such that $\p Q : t_1 \in \omega$. Then $F$ returns $F(\p Q :
t_1 \in \omega) + 1$. 

To show that $F$ terminates for all its arguments, consider the
sequence $t, t_1, t_2, {\ldots} $ of
terms  obtained throughout the execution of $F$.
We have \iizfr $\p t \in \omega$, \iizfr $\p t = S(t_1)$, \iizfr $\p t_1 = S(t_2)$
and so on. The length of the sequence is therefore exactly the natural
number denoted by $t$. 

\begin{corollary}[Numerical Existence Property]
If \iizfr $\p \exists x \in \omega.\ \phi(x)$, then there is a natural number
$n$ and term $t$ such that \iizfr $\p \phi(t) \land t = \ov{n}$. 
\end{corollary}
\proof
As before, use the Curry-Howard isomorphism to get a value $[t, M]$ such that $\p 
[t, M] : \exists x.\ x \in \omega \land \phi(x)$. Thus $\p M : t \in \omega
\land \phi(t)$, so $M \downarrow <M_1, M_2>$ and $\p M_1 : t \in \omega$.
Take $n = F(\p M_1 : t \in \omega)$. By patching together
the proofs \iizfr $\p t = S(t_1)$, \iizfr $\p t_1 = S(t_2)$, {\ldots}
,\iizfr $\p t_n = 0$ obtained throughout the execution of $F$, we get \iizfr $\p t = \ov{n}$.\qed

This version of NEP differs from the one usually found in the literature,
where in the end $\phi(\ov{n})$ is derived. However, \iizfr\ does not have the
Leibniz axiom for the final step. We conjecture that it is the only version
which holds in non-extensional set theories. More specifically, we
conjecture that there is a term $t$ and formula $\phi$ such that \iizfr $\p
\phi(t) \land t = \ov{n}$ and \iizfr\ does not prove $\phi(\ov{n})$. 

\section{Extensional \izfr}\label{lei}

We will show that we can extend our results to full \izfr. We work in \iizfr.

\begin{lemma}
Equality is an equivalence relation.
\end{lemma}
\proof
Straightforward.\qed

\begin{definition}
A set $C$ is \emph{L-stable}, if $A \in C$ and $A = B$ implies $B \in
C$. 
\end{definition}

Thus, L-stable sets are well-behaved as far as the atomic version of the
Leibniz axiom ($\forall a, b, c.\ a \in c \land a = b \to b \in c$) is concerned. 

\begin{definition}
A set $C$ is \emph{transitively L-stable} (we say that TLS(C) holds) if it is L-stable and every
element of $C$ is transitively L-stable. 
\end{definition}

This definition is formalized in a standard way, using transitive closure, available
in \iizfr, as shown e.g. in \cite{ar}. We denote the class of transitively L-stable sets
by $T$. The statement $V=T$ stands for $\forall A.\ TLS(A)$. The class $T$ in
\iizfr\ plays a similar role to the class of well-founded sets in ZF without
Foundation. 

\begin{lemma}
\izfr $\p V=T$. 
\end{lemma}
\proof
Straightforward $\in$-induction.\qed

The restriction of a formula $\phi$ to $T$, denoted by $\phi^T$, is defined
as usual, taking into account the following translation of terms:
\[
a^T \equiv a \quad \{ t, u \}^T \equiv \{ t^T, u^T \} \qquad \omega^T \equiv \omega \qquad
(\bigcup t)^T \equiv \bigcup t^T \qquad (P(t))^T \equiv P(t^T) \cap T 
\]
\[
(S_{\phi(a, \vec{f})}(u, \vec{u}))^T \equiv S_{\phi^T(a, \vec{f})}(u^T,
\overrightarrow{u^T}) \qquad 
(R_{\phi(a, b, \vec{f})}(t, \vec{u}))^T \equiv R_{b \in T \land \phi^T(a, b,
\vec{f})}(t^T, \overrightarrow{u^T})
\]
The notation $T \models \phi$ means that $\phi^T$ holds. 
\begin{lemma}
$T$ is transitive. 
\end{lemma}
\proof
Take any $A$ in $T$ and suppose $a \in A$. Then by the definition of $T$, $a
\in T$ as well.\qed

\begin{lemma}\label{t1}
If $A=C$ and $A \in T$, then $C \in T$. 
\end{lemma}
\proof
This is \emph{not} obvious, as there is no Leibniz axiom in the logic. 
Suppose $a \in C$ and $a = b$. Since $A=C$, $a \in A$. Since $A$ is L-stable, $b \in A$, so also $b \in C$. Thus $C$ is L-stable. 

If $a \in C$, then $a \in A$. Since $A \in T$ and $T$ is transitive, $a
\in T$. Thus $C$ is transitively L-stable.\qed

\begin{lemma}
Equality is absolute for $T$.
\end{lemma}
\proof
Take any $a, b \in T$. Suppose $(a = b)^T$. This means that for all $c \in T$,
$c \in a \iffl c \in b$. As $T$ is transitive, this is equivalent to for all
$c$, $c \in a \iffl c \in b$, so also $a = b$ in the real world. On the
other hand, if $\forall c.\ c \in a \iffl c \in b$, then obviously also
$\forall c \in T.\ c \in a \iffl c \in b$.\qed

The following three lemmas are essentially used to show that $T$ is closed
under the axioms of \izfr. 

\begin{lemma}\label{omegat}
$0 \in T$. If $A \in T$, then $S(A) \in T$. 
\end{lemma}
\proof
That $0 \in T$ is obvious. Take any $A \in T$. To show that $A \cup \{ A \}
\in T$, suppose $a \in A \cup \{ A \}$ and $a = b$. If $a \in A$, then by $A
\in T$ we have $b \in A$ and $a \in T$. If $a \in \{ A \}$, then $a = A$, so
also $b = A$ and by Lemma \ref{t1} $a \in T$. In both cases $b \in A \cup \{
A \}$ which shows the claim.\qed

The following two lemmas are proved together by mutual induction on the
definition of terms and formulas. 

\begin{lemma}\label{trieq}
For any term $t(a, \vec{f})$, $\forall a, b, \vec{f} \in T.\ (a = b \to t^T(a,
\vec{f}) = t^T(b, \vec{f})) \land t^T(a, \vec{f}) \in T$.
\end{lemma}
\proof
Case $t(a, \vec{f})$ of:
\begin{enumerate}[$\bullet$]
\item $a$, $f_i, \emptyset$. The claim is trivial.
\item $\omega$. It suffices to show that $\omega \in T$. We show by
$\in$-induction on $a$ that $\forall a.\ a \in \omega \to a \in T \land
\forall b.\ a = b \to b \in \omega$. Take any $a \in \omega$. 
Then either $a = 0$ or there is $y \in \omega$ such that $a = S(y)$. Take
any $b$ such that $a = b$. In the former case $b = 0$, so $b \in \omega$ and by Lemmas
\ref{t1} and \ref{omegat} we get $a \in T$. In the latter case, take this $y$. We have $b = S(y)$,
so $b \in \omega$. By $a = S(y)$, $y \in a$, so by the induction hypothesis
$y \in T$, thus by Lemma \ref{omegat} we also get $a \in T$. 
\item $\{ t_1(a, \vec{f}), t_2(a, \vec{f}) \}$. By the induction hypothesis, 
$t_1^T(a, \vec{f}) = t_1^T(b, \vec{f})$ and $t_2^T(a, \vec{f}) = t_2^T(b, \vec{f})$.
In order to show that $\{ t_1(a, \vec{f}), t_2(a, \vec{f}) \}^T =
\{ t_1(b, \vec{f}), t_2(b, \vec{f}) \}^T$, take any $A \in \{ t_1^T(a, \vec{f}), t_2^T(a, \vec{f}) \}$. Then
either $A = t_1^T(a, \vec{f})$ or $A = t_2^T(a, \vec{f})$, so either
$A = t_1^T(b, \vec{f})$ or $A = t_2^T(b, \vec{f})$, in both cases $A \in \{
t_1(b, \vec{f}) , t_2(b, \vec{f}) \}^T$. The other direction is symmetric and
we get $\{ t_1(a, \vec{f}), t_2(a, \vec{f}) \}^T = \{ t_1(b, \vec{f}), t_2(b, \vec{f}) \}^T$. 

Furthermore, by the induction hypothesis, $t_1^T(a, \vec{f}) \in T$ and
$t_2^T(a, \vec{f}) \in T$. Thus in both cases by Lemma \ref{t1}, $A \in T$. Suppose $A = B$.
Then either $B = t_1^T(a, \vec{f})$, or $B = t_2^T(a,
\vec{f})$. In both cases $B \in \{ t_1(a, \vec{f}), t_2(a, \vec{f}) \}^T$. Thus
we have shown that $\{ t_1(a, \vec{f}), t_2(a, \vec{f}) \}^T \in T$. 
\item $\bigcup t(a, \vec{f})$. Take any $A \in (\bigcup t(a, \vec{f}))^T
= \bigcup t^T(a, \vec{f})$. By the induction hypothesis, $t^T(a, \vec{f}) =
t^T(b, \vec{f})$. Thus there is $B \in t^T(a, \vec{f})$ such that
$A \in B$. Thus also $B \in t^T(b, \vec{f})$, so $A \in \bigcup t^T(b,
\vec{f})$. The other direction is symmetric and we get 
$(\bigcup t(a, \vec{f}))^T = (\bigcup t(b, \vec{f}))^T$. 

Furthermore, by the induction hypothesis, $t^T(a, \vec{f}) \in T$, 
so by transitivity of $T$, $B \in T$ and also $A \in T$. Finally, 
suppose that $C = A$. Then since $B \in T$, $C \in B$, so $C \in \bigcup t^T(a,
\vec{f})$. This shows the claim. 
\item $P(t(a, \vec{f}))$. By the induction hypothesis, $t^T(a, \vec{f}) =
t^T(b, \vec{f})$. Suppose $A \in (P(t(a, \vec{f})))^T$. Then $A \subseteq
t^T(a, \vec{f})$ and $A \in T$. Thus also $A \subseteq t^T(b, \vec{f})$, so $A
\in (P(t(b, \vec{f})))^T$. The other direction is symmetric and we get $
(P(t(a, \vec{f})))^T = (P(t(b, \vec{f})))^T$. 

Suppose $A = B$. Since $A \in T$, by Lemma \ref{t1} $B \in T$. 
It is easy to see that also $B \subseteq t^T(a, \vec{f})$, so $B \in P(t^T(a,
\vec{f})) \cap T = (P(t(a, \vec{f})))^T$.
\item $S_{\phi(a, \vec{f})}(t(a, \vec{f}), \overrightarrow{t(a, \vec{f})})$. Suppose $A \in
(S_{\phi(a, \vec{f})}(t(a, \vec{f}), \overrightarrow{t(a, \vec{f})}))^T$. Then
$A \in t^T(a, \vec{f}) \land \phi^T(A, \overrightarrow{t^T(a, \vec{f})})$.
By the induction hypothesis, $t^T(a, \vec{f}) \in T$ and
$\overrightarrow{t^T(a, \vec{f})} \in T$. Thus, by transitivity of $T$, $A \in T$.
Moreover, by the induction hypothesis, $t^T(a, \vec{f}) = t^T(b, \vec{f})$ and
$\overrightarrow{t^T(a, \vec{f})} = \overrightarrow{t^T(b, \vec{f})}$. Therefore $A \in t^T(b,
\vec{f})$. By Lemma \ref{tril} we get $\phi^T(A, \overrightarrow{t^T(b, \vec{f})})$. This shows
that $A \in (S_{\phi(a, \vec{f})}(t(b, \vec{f}), \overrightarrow{t(b, \vec{f})}))^T$. The
other direction is symmetric and we get $
(S_{\phi(a, \vec{f})}(t(a, \vec{f}), \overrightarrow{t(a, \vec{f})}))^T =
(S_{\phi(a, \vec{f})}(t(b, \vec{f}), \overrightarrow{t(b, \vec{f})}))^T$. 

Suppose $A = B$. By Lemma \ref{t1}, $B \in T$. Since $t^T(a, \vec{f}) \in T$, $B \in t^T(a, \vec{f})$. By
Lemma \ref{tril}, $\phi^T(B, \overrightarrow{t^T(a, \vec{f})})$ holds. Thus $(S_{\phi(a,
\vec{f})}(t(a, \vec{f}), \overrightarrow{t(a, \vec{f})}))^T \in T$. 
\item $R_{\phi(a, b, \vec{f})}(a, \vec{f})$. Suppose $A \in (R_{\phi(a, b,
\vec{f})}(t(a, \vec{f}), \overrightarrow{u(a, \vec{f})}))^T$ and $A = B$. This means that:
\begin{enumerate}[$-$]
\item $\forall x \in t^T(a, \vec{f}) \exists !y \in T.\ \phi^T(x, y, \overrightarrow{u^T(a, \vec{f})})$.
Take any $x \in t^T(b, \vec{f})$. By the induction hypothesis, $x \in t^T(a,
\vec{f})$. Thus there is $y \in T$ such that $\phi^T(x, y,
\overrightarrow{u^T(a, \vec{f})})$ and $\forall z \in T.\ \phi^T(x, z, \overrightarrow{u^T(a,
\vec{f})}) \to z = y$. We will now show that there is exactly one $y' \in T$
such that $\phi^T(x, y', \overrightarrow{u^T(b, \vec{f})})$. Take $y' = y$. 
By the induction hypothesis, $\overrightarrow{u^T(a, \vec{f})} = \overrightarrow{u^T(b, \vec{f})}$. 
By Lemma \ref{tril}, $\phi^T(x, y', \overrightarrow{u^T(b, \vec{f})})$.
Take any $z' \in T$ and assume $\phi^T(x, z', \overrightarrow{u^T(b, \vec{f})})$. By Lemma
\ref{tril}, $\phi(x, z', \overrightarrow{u^T(a, \vec{f})})$, so $z' = y'$. Thus we have
shown that $\forall x \in t^T(b, \vec{f}) \exists !y \in T.\ t^T(x, y,
\overrightarrow{u^T(b, \vec{f})})$. 
\item $\exists x \in t^T(a, \vec{f}).\ A \in T \land \phi^T(x, A, \overrightarrow{u^T(a, \vec{f})})$. Take this $x$.
By Lemma \ref{t1}, $B \in T$, so by Lemma \ref{tril}, $\phi^T(x, B, \overrightarrow{u^T(a, \vec{f})})$.  
Moreover, by Lemma \ref{tril}, $\phi^T(x, A, \overrightarrow{u^T(b, \vec{f})})$. 
Thus there is $x \in t^T(b, \vec{f})$ such that $\phi^T(x, A,
\overrightarrow{u^T(b, \vec{f})})$.
\end{enumerate}
Altogether, this shows that $A \in (R_{\phi(a, b, \vec{f})}(
t(b, \vec{f}), \overrightarrow{u(b, \vec{f})}))^T$. The other direction is symmetric and we get
$(R_{\phi(a, b, \vec{f})}(t(a, \vec{f}), \overrightarrow{u(a, \vec{f})}))^T =
(R_{\phi(a, b, \vec{f})}(t(b, \vec{f}), \overrightarrow{u(b, \vec{f})}))^T$. We have
also shown that $(R_{\phi(a, b, \vec{f})}(t(a, \vec{f}), \overrightarrow{u(a,
\vec{f})}))^T \in T$, so the proof is complete.\qed
\end{enumerate}

\begin{lemma}\label{tril}
$T \models L_{\phi(a, \vec{f})}$. In other words, $\forall a, b, \vec{f} \in
T.\ a = b \to \phi^T(a, \vec{f}) \to \phi^T(b, \vec{f})$. 
\end{lemma}
\proof
We show representative cases. Case $\phi$ of:
\begin{enumerate}[$\bullet$]
\item $t(a, \vec{f}) \in s(a, \vec{f})$ for some terms $t, s$. We need to show
that if $A, B, \vec{F} \in T$, $A = B$ and $t^T(A, \vec{F}) \in s^T(A,
\vec{F})$, then $t^T(B, \vec{F}) \in s^T(B, \vec{F})$. By Lemma \ref{trieq}, $t^T(A, \vec{F}) =
t^T(B, \vec{F}), s^T(A, \vec{F}) = s^T(B, \vec{F})$ and $s^T(A, \vec{F}) \in T$.
Therefore $t^T(B, \vec{F}) \in s^T(A, \vec{F})$, which entails $t^T(B, \vec{F})
\in s^T(B, \vec{F})$.
\item $\phi_1(a, \vec{f}) \to \phi_2(a, \vec{f})$. Take any $A,
B, \vec{F} \in T$, assume $A = B$,  $\phi^T_1(A, \vec{F}) \to
\phi^T_2(A, \vec{F})$ and $\phi^T_1(B, \vec{F})$. By the induction
hypothesis for $\phi_1$,  $\phi^T_1(A, \vec{F})$. Using the assumption we
obtain $\phi^T_2(A, \vec{F})$. By the induction hypothesis for $\phi_2$ we get 
$\phi^T_2(B, \vec{F})$. 
\item $\exists c.\ \phi_1(a, \vec{f}, c)$. Take any $A, B, \vec{F} \in T$,
assume $A = B$ and $\exists c \in T.\ \phi^T_1(A, \vec{F}, c)$. Then there is
a set $C \in T$ such that $\phi^T_1(A, \vec{F},C)$ holds. By the induction
hypothesis, merging $\vec{f}$ with $c$, we get $\phi^T_1(B, \vec{F}, C)$, so also $\exists
c.\ \phi^T_1(B, \vec{F}, c)$.\qed
\end{enumerate}

\begin{thm}\label{tlsmodel}
$T \models$\izfr. In other words, $T$ is an inner model of \izfr. 
\end{thm}
\proof
We proceed axiom by axiom. 
\begin{enumerate}[$\bullet$]
\item (EMPTY) Straightforward. 
\item (PAIR) Take any $A, B \in T$. That $\{ A, B \}$ satisfies the (PAIR) axiom in T 
follows by absoluteness of equality.
\item (UNION) Take any $A \in T$. Suppose $C \in \bigcup A$. Then there is
some $B$ such that $C \in B \in A$. Since $A$ is transitive, $B \in T$. On the
other hand, if there is $B \in T$ such that $C \in B \in A$, then obviously $C \in \bigcup A$. 
\item (INF) Suppose $C \in \omega$. Then either $C = 0$ or there is $y \in
\omega$ such that $C = S(y)$. We need to show that either $C = 0$ or there
is $y \in T$ such that $y \in \omega^T$ and $C = S^T(y)$. If $C = 0$, the claim
is trivial. Otherwise, suppose there is $y \in \omega$ such that $C = S(y)$.
Then $y \in C$, so by transitivity of $T$, $y \in T$. 
We also know that $\omega^T = \omega$ and $S^T(y) = S(y)$. The claim follows. 

On the other hand, suppose $C = 0$ or there is $y \in T$ such that $y \in
\omega$ and $C = S^T(y)$. In both cases, $C$ is trivially in $\omega$. 
\item (POWER) Take any $A, C \in T$. Suppose $C \in P^T(A)$. Then $\forall D
\in C.\ D \in A$, so also for all $D \in T$, $D \in C \to D \in A$. On the other hand,
suppose that for all $D \in T$, $D \in C \to D \in A$.  
To show that $C \in P^T(A)$, we need to show that $C \in T$ and for all $D \in C$, $D \in
A$. We already have the former. To show the latter, note that by
transitivity of $T$, any $D \in C$ is also in $T$, so by the assumption in
$A$. This shows the claim.
\item (SEP${}_{\phi(a, \vec{f})}$) Take any $A, \vec{F} \in T$ and suppose $C \in \{ x \in A\ |\ \phi(x,
\vec{F}) \}^T$. Then $C \in A$ and $\phi^T(C, \vec{F})$, which is what we need. 
On the other hand, if $C \in A$ and $\phi^T(C, \vec{F})$, then
also $C \in \{ x \in A\ |\ \phi^T(x, \vec{B}) \}= \{ x \in A\ |\ \phi(x, \vec{B}) \}^T$.
\item (REPL${}_{\phi(a, b, \vec{f})}$) Take any $A, \vec{F}, C \in T$ such that
$C \in \{ z\ | (\forall x \in A \exists! y.\ \phi(x, y, \vec{F})) \land 
\exists x \in A.\ \phi(x, z, \vec{F}) \}^T$. This is equivalent to 
$(\forall x \in A \exists! y.\ y \in T \land \phi^T(x, y, \vec{F})) 
\land \exists x \in A.\ C \in T \land \phi^T(x, C, \vec{F})$.
Since $A \in T$ and $T$ is closed under equality, it is also equivalent to $(\forall x \in T.\ x \in A \to
\exists y.\ y \in T \land \phi^T(x, y, \vec{F}) \land \forall z.\ z  \in T \to
z = y \to \phi^T(x, z, \vec{f})) \land \exists x \in T.\ x \in A \land C \in
T \land \phi^T(x, C, \vec{F})$, which is what we want. 
\item (IND${}_{\phi(a, \vec{f})}$) Take $\vec{F} \in T$ and suppose that $\forall x \in T. 
(\forall y \in T.\ y \in x \to \phi^T(y, \vec{F})) \to \phi^T(x, \vec{F})$. We have to
show that $\forall a.\ a \in T \to \phi^T(a, \vec{F})$.
We proceed by $\in$-induction on $a$. Take any $A \in T$. By the assumption
instantiated with $A$, $(\forall y \in T.\ y \in A \to \phi^T(y, \vec{F}))
\to \phi^T(A, \vec{F})$. We have to show that $\phi^T(A, \vec{F})$. It suffices to show 
that $\forall y \in T.\ y \in A \to \phi^T(y, \vec{F})$. Take any $y \in T
\cap A$. By the induction hypothesis for $y$, we get $\phi^T(y, \vec{F})$ and
the claim.
\item (L${}_{\phi(a, \vec{f})}$) Follows by Lemma \ref{tril}.\qed
\end{enumerate}

\begin{lemma}\label{tt}
For any term $t(\vec{a})$ and any formula $\phi(\vec{a})$, \izfr $\p \forall
\vec{a}.\ t^T(\vec{a}) = t(\vec{a}) \land \phi^T(\vec{a}) \iffl \phi(\vec{a})$.
\end{lemma}
\proof
By induction on the generation of terms and formulas. 
Case $t$ of:
\begin{enumerate}[$\bullet$]
\item $a, \omega, \emptyset$. The proof is obvious. 
\item $\{ t_1, t_2 \}$. By the induction hypothesis, $t_1^T = t_1$ and
$t_2^T = t_2$. So if $a \in \{ t_1^T, t_2^T \}$, then $a = t_1$ or $a = t_2$, so 
$a \in \{ t_1, t_2 \}$. The other direction is symmetric.
\item $\bigcup t_1$. By the induction hypothesis, $t_1^T = t_1$. If $a \in \bigcup t_1^T$, then there is
$b$ such that $a \in b \in t_1^T$, so $b \in t_1$ and $a \in \bigcup t_1$.
The other direction is symmetric.
\item $P(t_1)$. By the induction hypothesis, $t_1^T = t_1$. If $a \in P(t_1^T) \cap T$, 
then $a \subseteq t_1^T$, so also $a \subseteq t_1$ and consequently $a \in P(t_1)$. 
On the other hand, if $a \in
P(t_1)$, then by $V=T$ we also get $a \in T$, so $a \in (P(t_1))^T$. 
\item $\{ x \in t_1\ |\ \phi(x, \vec{u}) \}$. By the induction hypothesis,
$t_1^T = t_1, \vec{u}^T = \vec{u}$. Suppose $a \in \{ x \in t_1^T\ |\
\phi^T(x, \vec{u}^T) \}$. Then $a \in t_1^T$, so $a \in t_1$. Since 
$\phi^T(a, \vec{u}^T)$ and we work in \izfr, $\phi^T(a, \vec{u})$. By the
induction hypothesis, $\phi(a, \vec{u})$, so $a \in \{ x \in t_1\ |\ \phi(x,
\vec{u}) \}$. The other direction is symmetric. 
\item $\{ y\ |\ \forall x \in t_1\ \exists !y. \phi(x, y, \vec{u}) \land \exists
x \in t_1.\ \phi(x, y, \vec{u}) \}$. By the induction hypothesis, $t_1^T =
t_1$ and $\vec{u^T} = \vec{u}$. Suppose $a \in \{ y\ |\ \forall x \in t_1\ \exists !y. \phi(x,
y, \vec{u}) \land \exists x \in t_1.\ \phi(x, y, \vec{u}) \}^T$. Then:
\begin{enumerate}[$-$]
\item For all $x \in t_1^T$ there is exactly one $y \in T$ such that $\phi^T(x,
y, \vec{u}^T)$. By the induction hypothesis and $V=T$, we also have for all $x \in t_1$
there is exactly one $y$ such that $\phi(x, y, \vec{u})$.  
\item There is $x \in t_1^T$ such that $a \in T$ and $\phi^T(x, a,
\vec{u}^T)$. Then also there is $x \in t_1$ such that $\phi(x, a, \vec{u})$.
\end{enumerate}
Altogether, $a \in \{ y\ |\ \forall x \in t_1\ \exists !y. \phi(x, y, \vec{u})
\land \exists x \in t_1.\ \phi(x, y, \vec{u}) \}$. The other direction is
similar. 
\end{enumerate}
For the formulas, we show representative cases. Case $\phi$ of:
\begin{enumerate}[$\bullet$]
\item $t \in s$. By the induction hypothesis, $t^T =
t$ and $s^T = s$, so by the Leibniz axiom $t^T \in s^T$ is equivalent to $t \in s$.
\item $\forall a.\ \phi_1$. Suppose $\forall a.\ \phi_1$, then since $V = T$ we
have $\forall a \in T.\ \phi_1$. By the induction hypothesis, $\forall a \in
T.\ \phi_1^T$. The other direction is similar.\qed
\end{enumerate}

\begin{lemma}\label{liff}
\izfr $\p \phi$ iff \iizf $\p \phi^T$. 
\end{lemma}
\proof
The left-to-right direction follows by Theorem \ref{tlsmodel}. For the
right-to-left direction, if \iizf $\p \phi^T$, then also \izfr $\p \phi^T$
and Lemma \ref{tt} shows the claim.\proof

\begin{corollary}\label{dpnep}
\izfr\ satisfies DP, NEP and  TEP. 
\end{corollary}
\proof
For DP, suppose \izfr $\p \phi \lor \psi$. By Lemma \ref{liff},
\iizf $\p \phi^T \lor \psi^T$. By DP for \iizf, either \iizf $\p \phi^T$ or
\iizf $\p \psi^T$. Using Lemma \ref{liff} again we get either \izfr $\p \phi$ or
\izfr $\p \psi$.

For NEP, suppose \izfr $\p \exists x.\ x \in \omega \land \phi(x)$. By Lemma
\ref{liff}, \iizf $\p \exists x.\ x \in T \land x \in \omega^{T}.\
\phi^{T}(x)$, so \iizf $\p \exists x \in \omega^T.\ x \in T
\land  \phi^{T}(x)$. Since $\omega^{T} = \omega$, using NEP for \iizf\ we get
a natural number $n$ such that \iizf $\p \exists x \in \omega.\ x \in T
\land \phi^{T}(x) \land x =
\ov{n}$, thus also \iizf $\p \exists x \in T.\ x \in \omega^T \land
\phi^T(x) \land x = \ov{n}$. By Lemma \ref{liff} and $\ov{n} = \ov{n}^T$, we get \izfr $\p
\exists x.\ \phi(x) \land x = \ov{n}$. By the Leibniz axiom, \izfr $\p \phi(\ov{n})$.

For TEP, suppose \izfr $\p \exists x.\ \phi(x)$. By Lemma \ref{liff}, \iizf
$\p \exists x \in T.\ \phi^T(x)$. By TEP for \iizf, there is a term $t$ such
that \iizf $\p \phi^T(t)$. This implies \izfr $\p \phi^T(t)$. By Lemma
\ref{tt}, $t^T = t$, so by the Leibniz axiom in \izfr\ we get \izfr $\p \phi^T(t^T)$. 
Since $\phi^T(t^T) = \phi(t)^T$, by Lemma \ref{tt} we get \izfr $\p \phi(t)$.\qed

\begin{corollary}[Set Existence Property]\label{sep}
If \izfr $\p \exists x.\ \phi(x)$ and $\phi(x)$ is term-free, then
there is a term-free formula $\psi(x)$ such that \izfr $\p \exists !x.\ \phi(x) \land \psi(x)$.
\end{corollary}
\proof
Take the closed $t$ from Term Existence Property, so that \izfr $\p
\phi(t)$. By Corollary \ref{tdef} there is a term-free formula $\psi(x)$ defining $t$, so that
\izfr $\p (\exists !x.\ \psi(x)) \land \psi(t)$. Then \izfr $\p \exists
!x.\ \phi(x) \land \psi(x)$ can be easily derived.\qed

A different technique to tackle the problem of Leibniz axiom, used by Friedman in \cite{friedmancons}, 
is to define new membership ($\in^*$) and equality ($\sim$) relations in an intensional universe from scratch, so that $(V,
\in^*, \sim)$ interprets his intuitionistic set theory along with 
Leibniz axiom. Our $T$, on the other hand, utilizes existing $\in, =$ relations.
We present an alternative normalization proof, where the method to
tackle Leibniz axiom is closer to Friedman's ideas, in \cite{jatrinac2006}.

\section{Related work}\label{others}

Several normalization results for impredicative constructive set theories
much weaker than \izfr\ exist. Bailin
\cite{bailin88} proved  strong normalization of a constructive set theory
without the induction and replacement axioms. Miquel 
interpreted a theory of similar strength in a Pure Type System
\cite{miquelpts}. In \cite{miquel} he also defined a strongly normalizing
lambda calculus with types based on $F\omega.2$,
capable of interpreting \izfc\ without the $\in$-induction axiom. This result was
later extended --- Dowek and Miquel \cite{dowek} interpreted a version of constructive
Zermelo set theory in a strongly normalizing deduction-modulo system.

Krivine \cite{krivine} defined realizability using lambda calculus for classical set theory conservative
over ZF. The types for the calculus were defined. However, it seems that the types
correspond more to the truth in the realizability model than to provable
statements in the theory. Moreover, the calculus does not even weakly normalize.

The standard metamathematical properties of theories related to \izfr\ are well investigated.
Myhill \cite{myhill72} showed DP, NEP, SEP and TEP for IZF with Replacement and
non-recursive list of set terms. Friedman and \^S\^cedrov \cite{frsce1} showed SEP and
TEP for an extension of that theory with countable choice
axioms. Recently DP and NEP were shown for \izfc\ extended with various choice principles by Rathjen \cite{rathjenizf}.
However, the technique does not seem to be strong enough to provide TEP and SEP.

In \cite{jatrinac2006}, we show normalization of \izfr\ extended with
$\omega$-many inaccessible sets.

\section*{Acknowledgments}

I would like to thank my advisor, Bob Constable, for support and for
giving me the idea for $\li$ and this research, Richard Shore for
helpful discussions, David Martin for commenting on my ideas, Daria
Walukiewicz-Chrz{\fontencoding{T1}\selectfont\k a}szcz for the
higher-order rewriting counterexample, thanks to which I could prove
Theorem \ref{notweakly} and anonymous referees for helpful comments.
\bibliographystyle{alpha} \bibliography{latex8}

\end{document}